\documentclass[12pt,preprint]{aastex}
\makeatletter
\newcommand{\thereaction}{\thechapter-\Roman{reaction}}
\@addtoreset{reaction}{chapter}
\newcounter{reaction}
\newdimen\reactionindent
\newdimen\reactionarrspc
\def\@reactionnum{\hbox{\reset@font\rm(\thereaction)}}
\newcommand\chem[1]{$\rm #1$}
{%
  \@beginparpenalty\predisplaypenalty%
  \@endparpenalty\postdisplaypenalty%
  \refstepcounter{reaction}%
  \trivlist \item[]\leavevmode%
    \hb@xt@\linewidth\bgroup $\m@th
    \displaystyle%
    \hskip\reactionindent%
    \rm
}{%
    $\hfil 
    \displaywidth\linewidth\hbox{\@reactionnum}%
  \egroup%
  \endtrivlist%
}
\begin{document}
\newcommand{\hh}{H$_2$}
\newcommand{\hhh}{H$_3^+$}%
\newcommand{\hhd}{H$_2$D$^+$}
\newcommand{\hdd}{HD$_2^+$}
\newcommand{\ddd}{D$_3^+$}
\newcommand{\dd}{D$_2$}
\newcommand{\lrarrow}{$\leftrightarrow$}
\newcommand{\x}{$\times$}
\bibliographystyle{apj}
\title{The chemistry of multiply deuterated molecules in 
protoplanetary disks.  I. The outer disk}
\author{K. Willacy}
\affil{Jet Propulsion Laboratory, California Institute of Technology, MS 169--506, Pasadena CA 91109}
\email{Karen.Willacy@jpl.nasa.gov}
\begin{abstract}
We present new models of the deuterium chemistry in protoplanetary
disks, including, for the first time, multiply deuterated species. We
use these models to explore whether observations in combination with
models can give us clues as to which desorption processes occur in
disks. We find, in common with other authors, that photodesorption can
allow strongly bound molecules such as HDO to exist in the gas phase
in a layer above the midplane. Models including this process give the
best agreement with the observations. In the midplane, cosmic ray
heating can desorb weakly bound molecules such as CO and N$_2$. We
find the observations suggest that N$_2$ is gaseous in this region,
but that CO must be retained on the grains to account for the observed
DCO$^+$/HCO$^+$. This could be achieved by CO having a higher binding
energy than N$_2$ (as may be the case when these molecules are
accreted onto water ice) or by a smaller cosmic ray desorption rate
for CO than assumed here, as suggested by recent theoretical work.

For gaseous molecules the calculated deuteration can be
greatly changed by chemical processing in the disk from the input
molecular cloud values. On the grains singly deuterated species tend
to retain the D/H ratio set in the molecular cloud, whereas multiply
deuterated species are more affected by the disk
chemistry. Consequently the D/H ratios observed in comets may be
partly set in the parent cloud and partly in the disk, depending on
the molecule.

\end{abstract}


\section{Introduction}

Observations of deuterated molecules have long
been used to trace the physics and chemistry of interstellar clouds.  
The elemental abundance of deuterium relative to hydrogen is 
$\sim$ 10$^{-5}$, but the relative abundances of deuterated
to non--deuterated molecules can be much higher, especially in 
cold regions.  Deuterium is transferred from ions such as H$_2$D$^+$
and CH$_2$D$^+$ via gas phase ion-molecule reactions.
The reverse reactions, removing the deuterium, 
are inhibited at low temperatures,
due to the difference in the zero point energy of the deuterated
species to its non--deuterated equivalent.  
Another route to high levels of deuteration is via reactions
between species accreted on to the surfaces of dust grains.  These processes can lead to 
molecular D/H ratios of $>$ 0.01 in cold molecular clouds \citep[e.g.][]{turner01}.  
High levels of
deuteration are also seen in hot cores, protostellar disks and 
comets.  

Recently, multiply deuterated
molecules have been observed for the first time with high
levels of deuteration, e.g.\
NHD$_2$/NH$_3$ ranges from $\sim$ 0.005 (in the cold core, L134N) 
to $\sim$ 0.03 (in the low mass protostar 16293E) 
\citep{roueff00,loinard01},  ND$_3$/NH$_3$
$\sim$ 10$^{-3}$ \citep[in NGC133 IRAS4A and Barnard 1 -- ][]{lis02,tak02},
and D$_2$CO/H$_2$CO $\sim$ 0.01 - 0.4 
\citep[in cold cores and low mass protostars -- ][]{loinard02,bacmann03}.  
High abundances of deuterated methanol and formaldehyde have
been seen in star forming regions \citep{parise02,loinard02,bacmann03} and are attributed to 
the thermal desorption of grain
mantles accumulated during an earlier low temperature
phase, demonstrating the importance of grain surface reactions
in determining deuteration.  
In order to achieve such high molecular deuteration ratios, very high
atomic D abundances are required during the period when
ice mantles can accumulate, with D/H $\sim$ 0.2 - 0.3 \citep{caselli02,
parise02}.  
The recent models of \citet{roberts03} have shown that the inclusion
of HD$_2^+$ and D$_3^+$ can drive up the atomic D/H ratio to a level
where grain surface reactions can account for the observations of multiply
deuterated H$_2$CO and CH$_3$OH.  

A few deuterated molecules have now been observed in protostellar disks.
\citet{vd03} observed DCO$^+$ 
in TW Hya with DCO$^+$/HCO$^+$ = 0.035. 
This molecule was also observed in DM Tau, with DCO$^+$/HCO$^+$ 
= 4 $\times$ 10$^{-3}$ \citep{guilloteau06}.  \citet{cec04} observed H$_2$D$^+$
in TW Hya and DM Tau and found fractional abundances of 4 \x 10$^{-10}$ 
and 7 \x 10$^{-10}$ respectively.  They argued that the emission arises
in the midplane, where few other molecules exist because of
efficient freezeout, and thus observations of H$_2$D$^+$ provide
a means of determining the 
ionization fraction in this region.  \cite{cec05} presented
observations of HDO in DM Tau with a fractional abundance of $\sim$ 3 $\times$ 10$^{-9}$.
The emission arises in a region with temperatures below 25 K,
where HDO is expected to be frozen onto dust grains.  The presence
of HDO emission shows that an efficient non--thermal desorption process
must be acting, which \cite{dom05} suggest to be photodesorption.  
The detection of HDO and H$_2$D$^+$ in DM Tau has
been disputed by \cite{guilloteau06}.

The similarity between the molecular D/H ratios observed in comets and
those seen in molecular clouds is one reason why it has been suggested
that comets may be composed of interstellar material,
e.g.\ \citet{meier98a,meier98b,bm98,eberhardt95}.  This would mean that
the molecules had experienced little or no chemical processing in the
protosolar disk, yet models of deuterium chemistry in disks,
e.g.\ \citet{ah99a,ah01,aikawa02} show that extensive chemical
processing can occur, resulting in D/H ratios that are very different
from the ones set in the parent molecular cloud.  These disk models
have considered the chemistry of singly deuterated species only and
did not include grain chemistry, thereby ignoring a potentially
important process for determining the D/H ratios.  The model of
\citet{ah99a} considered the evolution of the chemistry in a cold cloud
core, through the infall stage and into the protostellar disk.  They
found, in the region of the disk where comets are expected to form,
that the molecular deuteration is affected by chemical processing in
the disk.  Therefore the abundances of molecules in comets does not
just reflect interstellar cloud abundances.  \citet{ah01} considered
only the disk chemistry, but used a disk model that was isothermal at
any given radius.  This was improved upon by \citet{aikawa02} who used
temperature and density distributions derived from the hydrodynamical
models of \citet{dalessio99}.  \citet{ah01} also ignored grain
reactions, but they found good agreement with the observed DCN/HCN and
DCO$^+$/HCO$^+$ ratios ($\sim$ 0.01).
 
Models have shown that 
in highly depleted regions, such as the disk midplane,
the deuteration of H$_3^+$ is efficient
and D$_3^+$ can become the dominant isotopomer \citep{cd05}.
This has 
consequences for the deuteration of other molecules since they
form from reactions of H$_3^+$, H$_2$D$^+$, HD$_2^+$ and D$_3^+$. 
\citet{roberts02,roberts03} demonstrated that the molecular D/H ratios
in dense molecular clouds can be significantly enhanced by  the inclusion
of HD$_2^+$ and D$_3^+$ in the models.  This effect is greatest 
in regions where molecules are heavily depleted by freezeout onto grains.
Since disks are very dense and therefore experience a high degree of
depletion, especially in the midplane, it seems likely that HD$_2^+$ and
D$_3^+$ could have a significant effect on the chemistry.

Given the importance of deuterium in tracing the origin of solar system
bodies and for understanding the thermal history of star forming regions,
together with the evidence that multiply deuterated forms of H$_3^+$ are important
for determining the molecular deuteration and the suggestion
that grain surface reactions may be important in determining the
D/H ratios, it seems timely to take another look at the
chemistry of deuterium in protoplanetary disks.  
Here we present
a model of the chemistry of a T Tauri disk that includes multiply
deuterated species, combined with grain surface chemistry.  We
consider how different desorption processes can affect the column 
densities and compare our results with observations with the
aim of better understanding the processes that are at work in disks.
We also use our models to consider how processing in disks may
affect the molecular D/H ratios observed in comets.

\section{The models}

In this section we discuss the processes that are included
in our models.  We wish to investigate how different
assumptions about the chemistry, in particular about desorption,
can affect the calculated abundances.  To do this we have
chosen one particular physical model to provide the density
and temperature distributions.  
We consider three desorption processes, thermal desorption, 
desorption due to cosmic ray heating of grains (CRH) and photodesorption.
All models include thermal desorption.  Model A excludes
non--thermal desorption,
Model B includes CRH,
Model C includes CRH and photodesorption, and Model D includes photodesorption but
not CRH.  Our models assume
that the disk is static, and that no mixing processes are acting.

\subsection{The chemical model}

Our basic chemical model is taken from the UMIST database RATE95 
\citep{rate95}.  The deuterated network is derived from this database using
the techniques described in \citet{millar89} and \citet{rm96}.
In the gas phase formation of  mono--deuterated species the deuterium is transferred
by reaction of H$_2$D$^+$, CH$_2$D$^+$ and C$_2$HD$^+$.  The rates for these
reactions are taken from \citet{rm00b}.  For multiply deuterated species
the reactions of HD$_2^+$, D$_3^+$, CHD$_2^+$, CD$_3^+$ and C$_2$HD$^+$
are important.
Rates for these reactions are taken from \citet{roberts04}.  Other reactions
involving deuterated species are assumed to have the same rates as their
non--deuterated counterparts.
For reactions involving complex species, the branching ratios 
must be determined.  In the absence of any information on
the reaction mechanism, this is done on a statistical basis, with the exception of reactions
where the same functional group appears as both a reactant and a product.  In
this case it is assumed that this group is preserved through the reaction
\citep{rm96}.  For example, in the dissociative recombination of 
\chem{CH_2DOH^+} only \chem{CH_2DOH} is formed and not \chem{CH_3OD}.

Our reaction network links 227 gas phase species (of which 115 are deuterated)
and 91 grain species (44 deuterated) by 9489 reactions.  
In order to reduce the time taken for the model to run, 
not every possible multiply deuterated species is included e.g.\ 
multiply deuterated forms of methanol are not included.

\subsection{The recombination of N$_2$H$^+$}

In RATE95 it is assumed that \chem{N_2H^+} recombines
with electrons to form N$_2$ and H only.  However, recent laboratory work
\citep{geppert04} has shown that this is in fact a minor
reaction pathway and instead
\begin{equation}
\label{eq:n2h+}
\hbox{\chem{N_2H^+}} + e^- \longrightarrow \hbox{NH} + \hbox{N}
\end{equation}
is the main route.  This result has consequences for the chemistry.
\citet{roberts03} found that the new branching ratios do not
affect the calculated abundances in gas phase only models, but do
have significant effects in models that include freezeout.
At low temperatures when freezeout is included, the destruction of \chem{N_2H^+} by
reaction with CO becomes less significant as the CO is depleted,
and so Reaction~\ref{eq:n2h+} gains in importance.
The formation of NH from reaction~\ref{eq:n2h+} results in the
removal of nitrogen from the gas in cold regions, since any NH that accretes onto
the grains will be rapidly hydrogenated to form NH$_3$, a molecule with a high binding
energy that is therefore not easily desorbed, unlike
N$_2$ which is very volatile.  Consequently, the new reaction pathways can
reduce the abundance of nitrogen--bearing molecules in 
the gas.  This reduction is less significant in regions where the grain
temperature is higher than about 20 K, where NH can be thermally desorbed
before reacting, and hence the formation of NH$_3$ ice is less efficient.

In disks, \citet{cd05} showed that 
N$_2$ is an important means of controlling the degree of deuteration. 
The presence of N$_2$
will prevent the transfer of deuteration along the chain of
isotopes from \chem{H_3^+} to \chem{D_3^+} and reduce the 
abundance of \chem{D_3^+} by destroying the less deuterated
isotopomers before they have a chance to form \chem{D_3^+}.  
The new reaction pathway means that the abundance of N$_2$ will
be reduced in the midplane, and hence the degree of deuteration of
\chem{H_3^+} will be higher.

\subsection{Freezeout and desorption processes}

Given the high densities and cold temperatures found in much of the
outer disk, molecules that hit a grain are likely to stick to it
efficiently.  We assume that all species freezeout
at the same rate with a temperature independent
sticking coefficient of 0.3.  For positive ions the freezeout rate
is increased slightly, since grains are likely to have a negative
charge \citep{un80} and therefore there could be a stronger attraction
between the positive ions and the negatively charged
grains.  The freezeout rate of positive ions is assumed to increase
by a factor $C$ = $1 + 16.71 \times 10^{-4}/(a T_{gr})$ where
$a$ is the grain radius and $T_{gr}$ is the grain temperature
\citep{un80}.  Ions are assumed to recombine on the grain surfaces
in the same way that they do when reacting with electrons in the
gas, e.g.\ in the gas phase
\begin{equation}
\hbox{HCO$^+$} + e \longrightarrow \hbox{H} + \hbox{CO}
\end{equation}
and on the grain
\begin{equation}
\hbox{HCO$^+$} + gr \longrightarrow \hbox{H}:gr + \hbox{CO}:gr
\end{equation}
where H$:gr$ and CO$:gr$ represent H and CO on the grains.
The exceptions are \chem{N_2H^+} and \chem{N_2D^+} where we
follow \citet{roberts04} in assuming that these ions form N$_2$
rather than NH and ND when they freezeout.
All species are assumed to freezeout, except for He which 
has a very low binding energy and is therefore 
easily thermally desorbed even at very low temperatures.
Any He$^+$ that hits a grain is neutralized and returned
immediately to the gas. 

We include thermal desorption and desorption due to cosmic
ray heating of grains.  Cosmic ray heating is able to maintain a 
low level of some
volatile molecules such as CO and N$_2$ in the cold midplane
of our models.  These molecules can destroy ions such as H$_3^+$
and its deuterated isotopes, and produce ions such as HCO$^+$
and DCO$^+$ in the midplane and thereby affect the 
ionization level in this region.  We have therefore run models without
cosmic ray heating to determine how the inclusion of this process
will affect molecular abundances and the ionization level in cold
regions of the disk.  The rates for cosmic ray heating are taken
from \citet{hh93}, using updated binding energies for some species,
notably CO, for which we use the value determined by \citet{oberg05}.
Table~\ref{tab:be} gives the binding energies ($E_D$) used in our models.

Thermal desorption rates are calculated from
\begin{equation}
k_{therm} = \nu_0 e^{-E_D/T_{gr}} \mbox{~~~s$^{-1}$}
\end{equation}
where $T_{gr}$ is the grain temperature,  and
$\nu_0$ is the frequency of oscillation between
the absorbate and the surface given by 
\begin{equation}
\label{eq:nu0}
\nu_0 = \sqrt (2 n_s E_D/ \pi^2 m)
\end{equation}
where $n_s$ is the surface density of sites ($\sim$ 1.5 $\times$ 10$^{15}$ cm$^{-2}$) and
$m$ is the mass of the accreting species.

\subsection{Photodesorption}

\citet{cec05} recently observed a high column density of
HDO in the disk around DM Tau.  They found N(HDO) $\sim$ 1.6 $\times$ 10$^{13}$
cm$^{-2}$ in the outer disk, where the density is $\sim$ 10$^6$ cm$^{-3}$ and 
the temperature is $<$ 25 K.  The emission comes from above the midplane
and corresponds to a relatively high fractional abundance of $\sim$ 3 $\times$ 
10$^{-9}$.  At these densities and temperatures, water is expected to be
completely removed from the gas by accretion onto grains, where it will
remain in the absence of a non--thermal desorption process.  Cosmic
ray heating is not efficient enough to remove such a strongly bound molecule
as water. \citet{dom05} suggested that photodesorption, arising from the action
of the interstellar far--UV field could be efficient enough to retain water vapor
in the gas phase in the cold outer disk, and so account for the observed column
densities.  They found that photodesorption can maintain a layer of water
vapor above the midplane with a fractional abundance of $\sim$ 3 $\times$ 10$^{-7}$.
The average calculated value of N(H$_2$O) in their model was $\sim$1.6 $\times$
10$^{15}$ cm$^{-2}$, and using this combined with  the observed N(HDO) they
deduce HDO/H$_2$O = 0.01 in DM Tau.

We have previously investigated the effects of photodesorption on the chemistry
of disks \citep{wl00} and found that it can retain high abundances of many molecules in
a layer above the midplane, even in cold disks.  Here we revisit the idea of
photodesorption and investigate not only how it affects the
abundances but also how it impacts the deuteration of molecules (Models C and D)

The rate of 
photodesorption is given by
\begin{equation}
k_{pd} = G_0 Y e^{-1.8 A_V} \pi a^2 n_g \mbox{~~~s$^{-1}$}
\end{equation}
where $G_0$ is the radiation field in units of the Habing field
(10$^8$ photons cm$^{-2}$s$^{-1}$), $A_V$ is the visual
extinction and $Y$ is the photodesorption 
yield.  We use the temperature dependent value of $Y$ as determined
experimentally by \citet{westley95}; From Figure 2 of
\citeauthor{westley95}\ Y = 0.003 molecules per photon
at $T$ $<$ 50 K and rises to 0.075 at $T$ = 100 K.

We include the effects of desorption caused by photons
from the interstellar radiation field, the stellar field and
from the cosmic ray induced photon field \citep{pt83}.

\subsection{Grain surface chemistry}

Previous models of deuterium chemistry in disks have
ignored grain surface reactions
and thereby have excluded a potentially very important
contribution to the molecular D/H ratios.  The inclusion of grain 
chemistry is problematic, as the simplest way is to 
use the rate equation method, but this gives very different
surface abundances than the more exact Monte Carlo model.
\citet{caselli02}
found that using the H--atom scan rates of \citet{katz99},
which are much lower than usually assumed, produces grain mantle
abundances in better agreement with Monte Carlo models than
the results from standard rate equation models.  The 
work of Katz ETA al.\ has been disputed, e.g.\ by 
\citet{horn03} who find that H$_2$ forms very efficiently on 
grains, possibly by H--atoms tunneling through barriers on the
grain surface.  However, because of the advantage in going some way to
correcting the short-comings of the rate equation method
we use the
Katz wt al.\ scan rates here, and assume that the scan rate
for D--atoms is also slow.

We assume that reactions on grain surfaces can occur only if one of the 
reactants is an atom.  All other species are assumed to be immobile.
The reaction set is taken from \citet{hh93} with 
the addition of the equivalent deuteration reactions.  
Activation barriers for the deuterium reactions are taken from
\citet{caselli02}.  The rates are temperature dependent 
and the reaction rate between two species, $i$ and $j$, is given by
\begin{equation}
k_{ij} = \kappa_{ij} (t_i^{-1} + t_j^{-1})/(N_s n_d)
\end{equation}
where 
$\kappa_{ij}$ = e$^{-E_a/kT_{gr}}$ ($E_a$ is the activation barrier
to the reaction), $N_s$ is the total number of sites on the
surface of the grain and $n_d$ is the grain number density.
The rate at which $i$ and $j$ scan the surface of the grain is
given by $t_i$ and $t_j$:
\begin{equation}
t_i^{-1} = \nu_0 e^{(-0.3 E_D/T_{gr})}
\end{equation}
where $E_D$ is the binding energy (from Table~\ref{tab:be})
and $\nu_0$ is given by Equation~\ref{eq:nu0}. 

Important parameters are the binding energies of H and D atoms
and their reaction rates on the grains.  For the formation rate
of H$_2$ we have used the work of \cite{ct02,ct04}
who developed a model of this process that fits the experimental
data for the reaction of two hydrogen atoms on silicate and amorphous carbon grains.
They find that
H$_2$ formation can be efficient at temperatures up to 500 K on
these surfaces.   This model takes
into account the possibility of both chemi- and physi-sorption
and assumes a high binding energy of 600 K for atomic hydrogen atoms
adsorbed onto a silicate surface.  This value of $E_D$(H) is rather
higher than the 350 K \cite{ta87} which we have assumed in
our previous work, and which results in a greatly reduced H$_2$ formation
rate at $T$ $>$ 15 K.
For deuterium atoms we follow \citet{caselli02} and
take $E_D$(D) = $E_D$(H) + 21 K (21 K is the
zero point energy difference between the hydrogen and deuterium atoms).
Recent calculations of the binding energy of H--atoms
adsorbed onto water ice have found values ranging from $\sim$ 400 K \citep{alh}
to $\sim$ 575 K \citep{perets05}.  Our choice of 600 K is therefore
a little high, but consistent with the larger of these calculated values.
Given the uncertainty in $E_D$(H) we compare 
the results for $E_D$(H) = 600 K with those for $E_D$(H) = 350 K in 
Section~\ref{sec:h_be} to determine the effects of our choice on the
chemistry.

\subsection{Ionization processes}

Ionization in the disk can arise from several sources:
\begin{enumerate}
\item {\it UV photons}  \\
In the surface layers ionization by UV photons from both
the stellar and interstellar radiation fields is important.
We assume that the stellar photons travel out radially
and that the interstellar photons hit the disk vertically.
The two fields are combined
to give an overall UV flux at each position in the disk.  The
strength of the stellar UV field for a T Tauri star 
has been estimated as 10$^{4}$ times
the interstellar radiation field ($G_0$) at 100 AU by \citet{hg96}.
More recent observations by \citet{bergin03} have 
estimated a much lower value of a few hundred times $G_0$.
We have chosen a value of 500 $G_0$ at 100 AU.  In common with 
\citet{aikawa02} we assume that the stellar field
does not dissociate CO and \hh, although these molecules are
dissociated by the interstellar radiation field.  We use
the approach of \citet{lee96} to describe the self--shielding
of both molecules.
The \citeauthor{lee96}\ model is based on a slab model
of a molecular cloud and provides data for the shielding due
to H$_2$, CO and dust as a function of column density.
The linewidths assumed are 3 kms$^{-1}$ -- much larger than observed in disks,
where the velocity dispersion in the outer disk is almost thermal \citep{gd98}.
To take account of this we have followed \cite{ah99} in scaling 
the column densities in Table 10 of \citeauthor{lee96}\ by $c_s$/3 kms$^{-1}$
where $c_s$ is the sound velocity.  This scaling factor is only required
for H$_2$ since CO dissociation lines are broader due to predissociation.

HD does not self--shield.  However some of its lines do
overlap with those of H$_2$ allowing shielding by the H$_2$ lines
to reduce its photodissociation rate.  \citet{barsuhn77} estimated
that the HD photodissociation rate will be reduced by 1/3, assuming that the 
overlapped HD lines are totally shielded by H$_2$.
We adopt this factor and scale it where necessary to take
account of the region where H$_2$ does not fully self--shield.


\item {\it Cosmic rays}\\
Cosmic rays can produce ionization in the disk if the surface
density is low enough for them to penetrate i.e.\ less that
150 g cm$^{-2}$ \citep{un81}.  This
is always the case in the region of the disk considered here.
Cosmic rays can cause ionization both directly and indirectly 
by producing photons from their interaction with H$_2$.
The rates of these processes are taken from the UMIST ratefile
with an assumed cosmic ray ionization rate of 1.3 $\times$ 10$^{-17}$ s$^{-1}$.

\item {\it Radioactive isotope decay}\\
The decay of radioactive nucleides is an additional source
of ionization. $^{26}$Al can decay to form excited $^{26}$Mg,
which in turn decays by either positron decay or by electron capture.
We include ionization due to these processes with a
rate:
\begin{equation}
\zeta_{Al} = 6.1 \times 10^{-18} \hbox{s$^{-1}$}
\end{equation}
\noindent \citep{un81}

\item {\it X--rays}\\
T Tauri stars are strong X--ray emitters.  \citet{ig99}
modeled the effects of X--ray ionization in a disk and found that
it can be effective in the surface layers where the attenuation length of
X--rays is very small.  \citet{ah01} included the effects
of X--rays in their model of the chemistry in the outer disk, 
but found that they did not significantly affect the results
for this region. Based on this we have chosen to ignore the
effects of X--rays in these models.

\end{enumerate}

\subsection{Input abundances for the disk model}

We assume that the material which is incorporated into
the disk has first been processed to some extent in the
parent molecular cloud.  We therefore use the
abundances produced by a molecular cloud model that has
been run for 1 Myrs as inputs to the disk model.  The elemental
abundances used as inputs to the cloud model are listed
in Table~\ref{tab:cloud}.  We assume that initially
all hydrogen is molecular and that the deuterium is 
in HD.  All other elements are in their atomic form,
with the exception of carbon which is ionic.  The
molecular cloud model is run at a density of 2 $\times$ 10$^4$
cm$^{-3}$, T = 10 K and a visual extinction of 10 magnitudes.
It includes all the processes that are included in the disk
model, with the exception of photodesorption and ionization by
the decay of radioactive nucleides.  The molecular abundances
which are used as inputs to the disk model are given in Table~\ref{tab:input}.
For comparison we also include the molecular abundances observed in 
the dark cloud TMC--1.  In general we find good agreement
between these observations and the abundances calculated in the
molecular cloud model.

\subsection{The disk model}

\begin{figure}
\caption{\label{fig:dalessio}The density and temperature distributions used
in this paper.  These are from the models of \citet{dalessio01}, with $\dot M$
= 10$^{-8}$ M$_\odot$ yr$^{-1}$, $M_*$ = 0.7 M$_\odot$, $T_*$ = 4000 K, $R_*$ = 
2.5 R$_\odot$. The grain size distribution is given by $n(a)$ $\propto$ $a^{-3.5}$ for
0.005 $<$ $a$ $<$ 0.25 $\mu$m, where $a$ is the grain radius.
The surface density $\Sigma_0$ = 
25 gcm$^{-2}$ at 10 AU and varies as $1/R$ for $R$ $>$ 10 AU.
The mass of the disk is 0.063 M$_\odot$ out to $R$ = 400 AU.}
\epsscale{1.5}
\plottwo{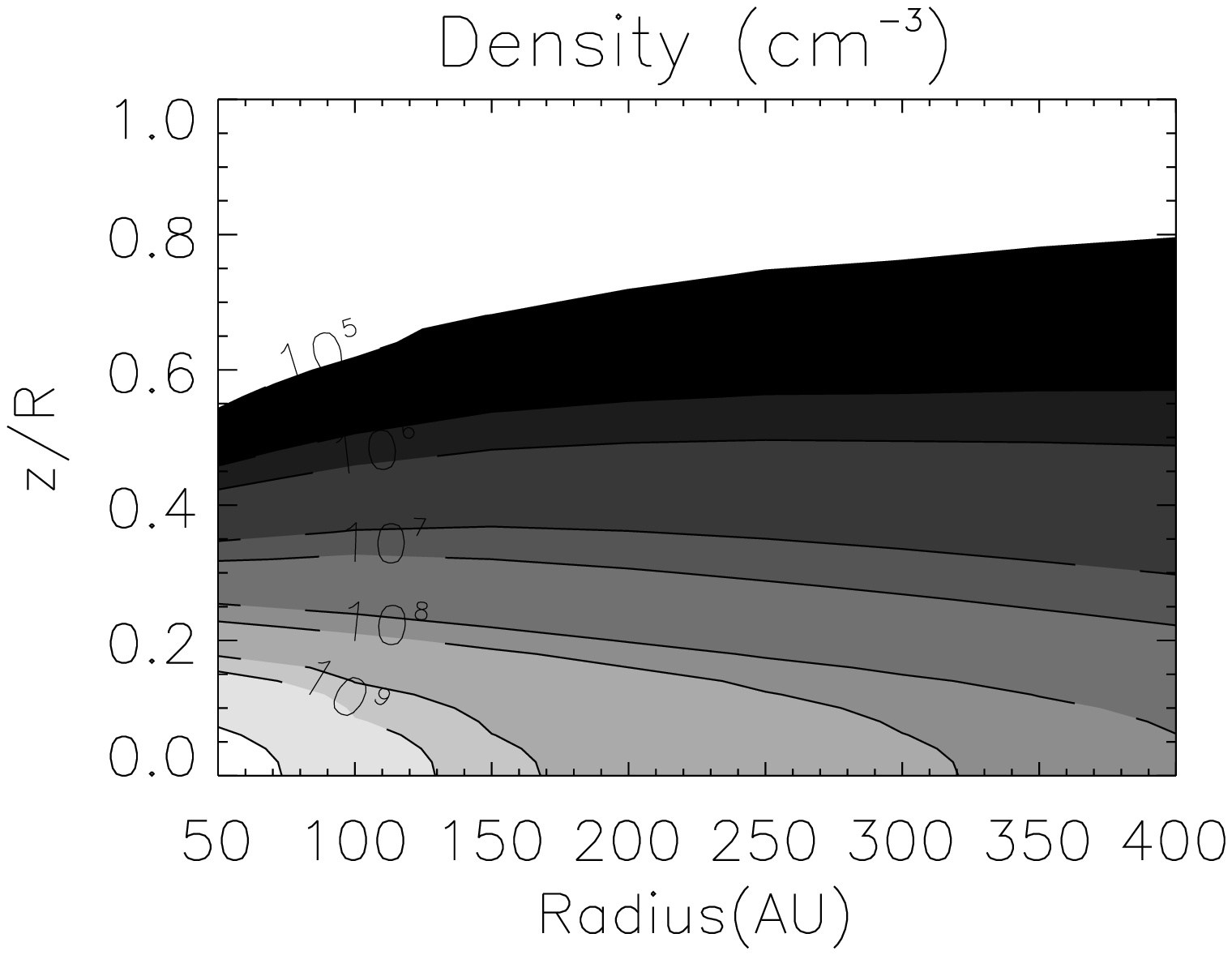}{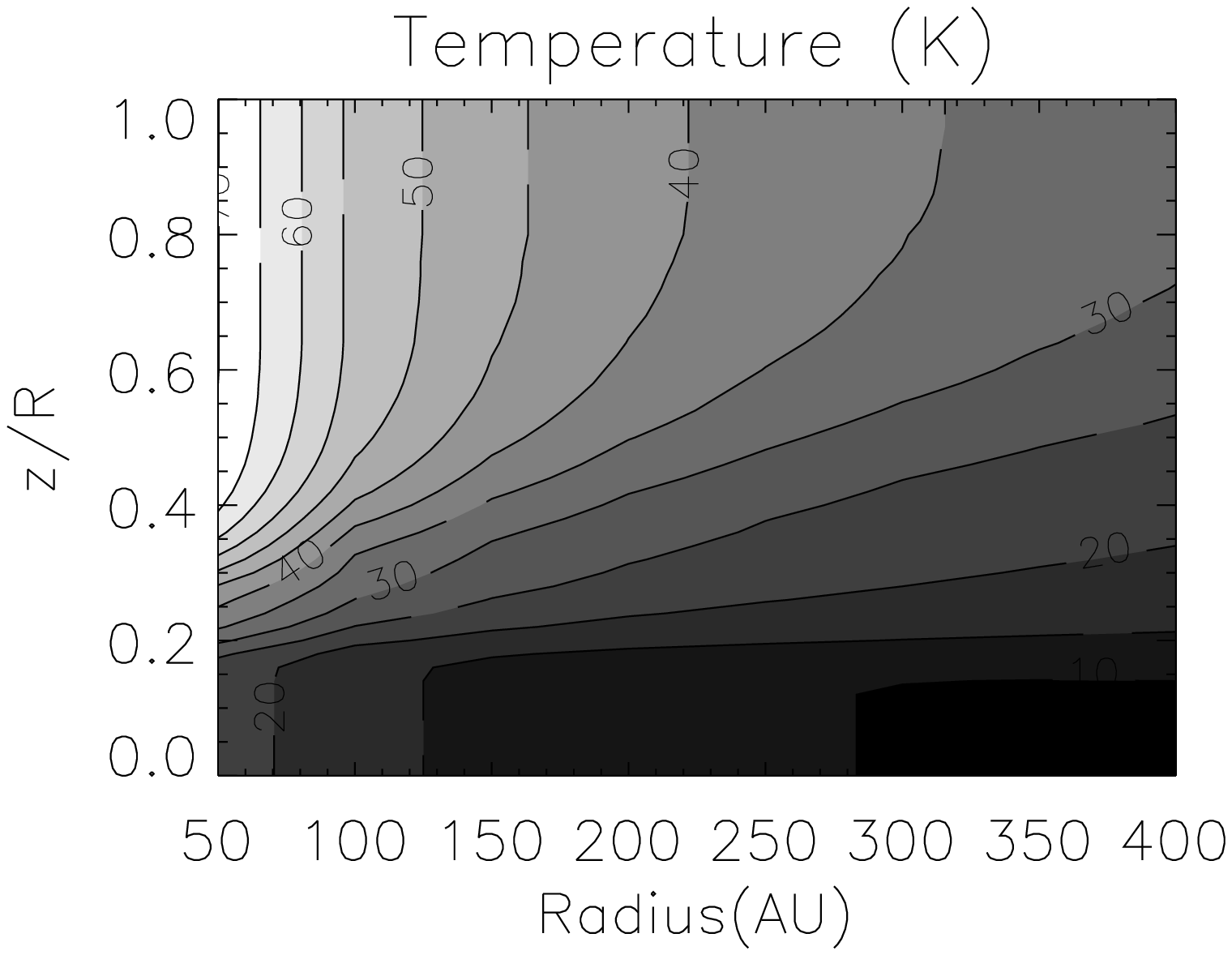}
\end{figure}

We have used the disk model of \citet{dalessio01} with 
a mass accretion rate, $\dot M$  = 
10$^{-8}$ M$_\odot$ yr$^{-1}$ to provide the physical parameters of the
disk used here.
The central star has a temperature of 4000 K, a mass of 0.7 M$_\odot$ and a radius of 2.5 R$_\odot$.  
The surface density is 25 gcm$^{-2}$ at 10 AU and varies as 1/$R$ for $R$ $>$ 10 AU.  
The disk mass out to 400 AU is 0.063 M$_\odot$.
The grain size distribution is given by $n(a)$ $\propto$ $a^{-3.5}$ with 
0.005 $\mu$m $\leq$ $a$ $\leq$ 0.25 $\mu$m.  
The density and temperature distributions are 
shown in Figure~\ref{fig:dalessio}.
The gas temperature is set equal to the grain temperature, and 
the density and temperature are assumed to remain constant over the
timescale considered in the chemical model.

\section{Results}

In the discussion that follows we look at the results of
our models.  Since the majority of molecules have been observed in the 
gas phase of disks we concentrate on these, but we also 
include a discussion of the grain chemistry and abundances
because in some regions of the disk the solid phase is the dominant
component, 
and because of the importance of grain chemistry in determining
the gas phase composition and molecular deuteration. We look
at how the molecular abundances vary with height above the midplane,
and at the radial column density distribution.  Finally,
we compare our results with the available observations and use
this to determine whether we can say anything about the nature of the
desorption processes acting in the disk.

\subsection{\label{sec:vert}The vertical molecular distribution}

\subsubsection{Gas phase}
We begin by looking at the vertical abundance distributions
at $R$ = 250 AU and a time of 1 Myrs. 
In this section we concentrate on Model B; differences in the chemistry
that arise because of desorption processes included in the 
other models are discussed in the next sections.

\begin{figure}
\caption{\label{fig:three}Schematic of the chemical structure of
a disk, showing how it can be divided into three layers.  In the
midplane the high densities and low temperatures ensure that most
molecules are frozen out onto the surfaces of the dust grains.
At the surface, the high UV field dissociates molecules into their
constituent atoms and ions.  Between these two lies a molecular layer
where the observed emission originates.}
\vspace*{0.5in}
\epsscale{1.0}
\plotone{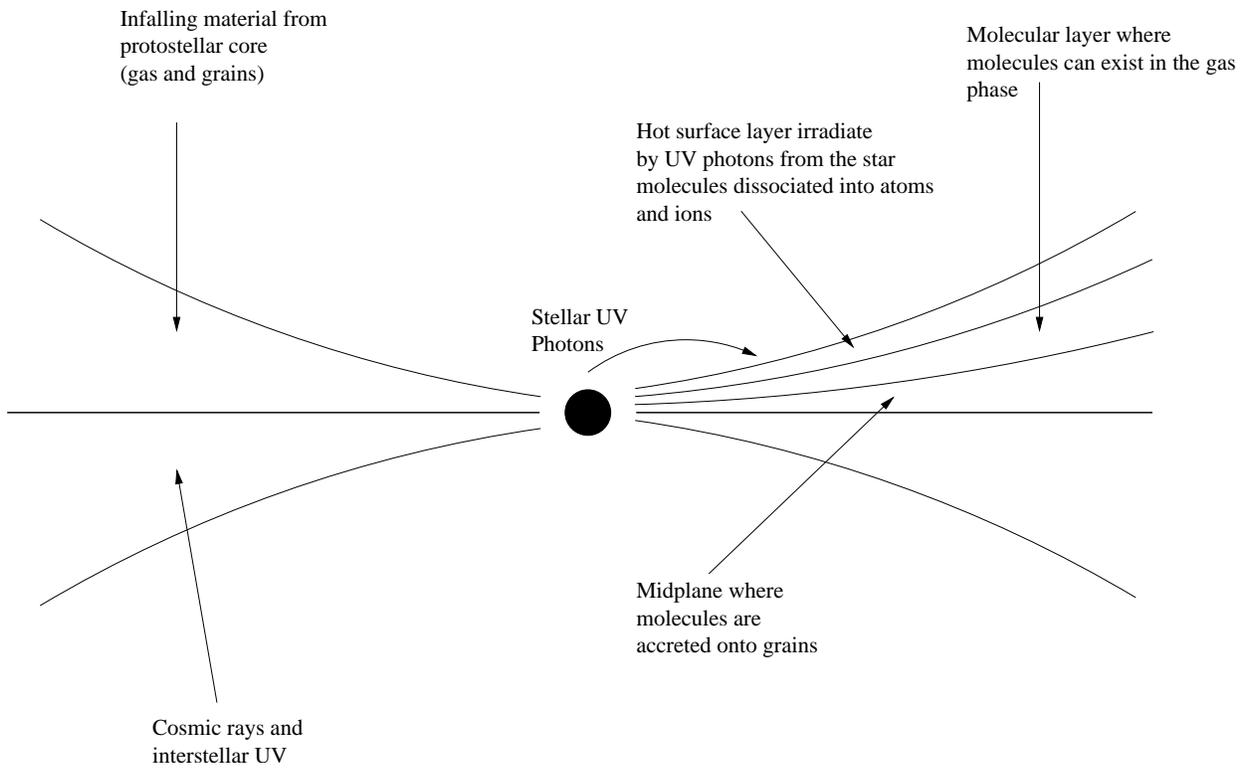}
\end{figure}

\begin{figure}
\caption{\label{fig:frac_250_a}The fractional abundances calculated at $R$ = 250 AU
at a time of 1 Myrs for Model B as a function of $z$.}
\epsscale{0.8}
\plotone{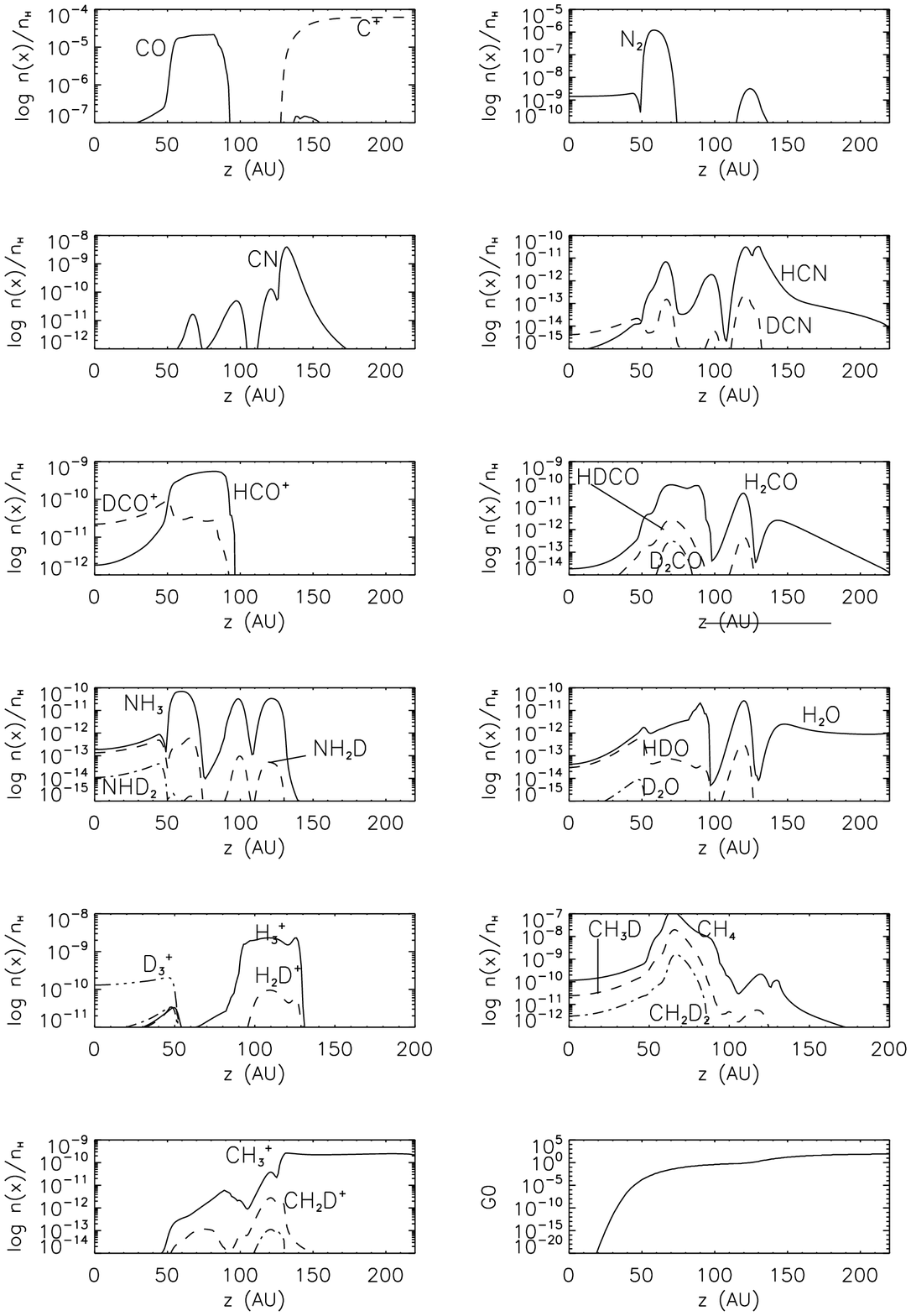}
\end{figure}

The fractional abundances
as a function of height, $z$, above the midplane
are displayed in Figure~\ref{fig:frac_250_a}.
The three layer structure (see Figure~\ref{fig:three})
found by previous authors \citep{aikawa02,wl00, ah99}
is clearly seen, with most molecules having low abundances in
the midplane (due to freezeout at $z$ $<$ 50 AU) and in the surface layers (due
to photodissociation at $z$ $>$ 100 AU), and abundance peaks in the molecular
layer, which for this model is between 50 and 95 AU above
the midplane. 

The midplane is not completely devoid of molecules since desorption by
CRH can maintain low levels of 
CO and N$_2$ in this region.  The reaction of these two species
with H$_3^+$ and its deuterated isotopomers results in 
the formation of N$_2$H$^+$, N$_2$D$^+$, HCO$^+$ and DCO$^+$.
Since the temperature is cold, there is a high level of deuteration
of H$_3^+$, producing high D/H ratios in the daughter molecules.
The presence of CO and N$_2$ in the gas reduces the deuteration of H$_3^+$
compared to Model A (which does not have non--thermal desorption) from
D$_3^+$/H$_3^+$ $\sim$ 70 to a value of $\sim$ 25.
Deuterated H$_3^+$ forms from the reaction 
of HD with the previous molecule in the chain e.g.\ H$_2$D$^+$ reacts with 
HD to form HD$_2^+$.  If CO and N$_2$ are present in the gas, 
H$_3^+$ and its deuterated isotopomers are more likely to react with
these molecules than with HD, reducing the formation rate of the deuterated
molecules (see \citet{cd05} for details).

The molecular layer begins at the point where thermal desorption
can remove some molecules, mainly CO, N$_2$ and CH$_4$ from the
grains.  The upper boundary is set by photodissociation.  
In the molecular layer most of the CO cycles between its
gaseous and solid phases, but a small proportion is
broken up by reaction with He$^+$ releasing C$^+$ and O, 
some of which go on to form molecules in the gas such as H$_2$O and H$_2$CO.
At $z$ $\sim$ 90 AU photodissociation can begin to destroy CO, forming
carbon and oxygen atoms which do not reform CO but instead are
converted into other molecules such
as H$_2$O and hydrocarbons.  These subsequently freezeout and are
not sufficiently volatile to be thermally desorbed. 
Carbon and oxygen are thus removed from the gas, leading to 
a sharp decrease in the abundance of CO at $z$ $\sim$ 90 AU.
At the very top of the disk most of the carbon is in its ionic
form, since it is efficiently produced by photoionization and 
the temperature is warm enough that its freezeout is not efficient.

N$_2$ exists in a narrower region than CO.  CRH ensures
that it is present in low abundances in the midplane.
Above this, just below the level at which
thermal desorption becomes efficient ($\sim$ 50 AU), its fractional abundance
shows a decrease from its midplane value.
The loss of N$_2$ is a result of its conversion
into N$_2$H$^+$ and N$_2$D$^+$.  These ions can be destroyed either
by reaction with CO or by recombination with electrons.
At $z$ $\sim$ 45 AU, the main process is recombination, which produces
mostly nitrogen atoms with NH and ND.  These freezeout and are quickly hydrogenated to form ices of
ammonia and its deuterated equivalents.  Once CO begins to thermally desorb, 
its reaction with N$_2$H$^+$ and N$_2$D$^+$ becomes their main destruction route, 
producing N$_2$, rather than N and NH or ND, and so less nitrogen is
lost by conversion into non--volatile ices.

Moving to higher $z$, there is a relatively narrow peak in the abundance
of N$_2$ due to thermal desorption.  This is
followed by a rapid decrease in abundance at $z$ $\sim$ 65 AU
where photodissociation becomes efficient 
and the molecule is destroyed.  The resulting nitrogen atoms
can freezeout onto grains where they react quickly with hydrogen atoms
to form NH, the first step towards forming ammonia ice.
However
NH is very volatile, with a binding energy of only 604 K \citep{ar77} and
so can be easily returned to the gas (if the temperature
is high enough) before hydrogenation can occur.
The volatility of NH drives the nitrogen chemistry in the gas at $z$
$>$ 100 AU.  Formation of gaseous molecules such as
HCN and CN can proceed via
\begin{equation}
\hbox{NH} \stackrel{\hbox{C$^+$}}{\textstyle \longrightarrow} \hbox{CN$^+$} \stackrel{\hbox{H$_2$}}{\textstyle \longrightarrow}
\hbox{HCN$^+$} \stackrel{\hbox{H$_2$}}{\textstyle \longrightarrow} \hbox{HCNH$^+$} \stackrel{\hbox{$e$}}{\textstyle \longrightarrow}
\hbox{HCN} \stackrel{\hbox{$h\nu$}}{\textstyle \longrightarrow} \hbox{CN}
\end{equation}

The photodissociation of CN produces nitrogen atoms and the cycle
starts again.  This process is sufficiently rapid to maintain low abundances
of HCN in the gas phase, even in regions with high UV fields.  

Several other molecules also have small abundance peaks at the
surface of the disk e.g.\ CH$_3^+$ and H$_2$O (Figure~\ref{fig:frac_250_a}).  
These are formed in the warm gas by efficient reactions involving the
atoms and ions produced by the photodissociation of molecules.  The
number of molecules produced is relatively small and does not greatly
affect the calculated column densities.

The effects of the increase in temperature with $z$ on the
D/H ratios can be seen in Figure~\ref{fig:frac_250_a}.
H$_3^+$ is highly deuterated at $z$ $<$ 50 AU, with the most abundant
isotopomer being D$_3^+$.  At higher $z$, H$_3^+$ is more
abundant and of the deuterated forms only H$_2$D$^+$ is important.
The peak in H$_3^+$ between $z$ = 90 and 130 AU is due to the
heavy depletion of molecules that would normally destroy 
it e.g.\ CO and N$_2$.  The variation in D/H ratio with $z$
for H$_3^+$ translates into variations in the D/H ratio
of other molecules e.g.\ DCO$^+$/HCO$^+$ $>$ 1 at $z$ $<$ 50 AU, 
but at $z$ $>$ 50 AU HCO$^+$ is the dominant isotopomer.

\subsubsection{How photodesorption affects
the calculated gas phase abundances and the molecular deuteration}

\begin{figure}
\caption{\label{fig:frac_250_c}The fractional abundances calculated at $R$ = 250 AU
at a time of 1 Myrs for Model C as a function of $z$.}
\epsscale{0.8}
\plotone{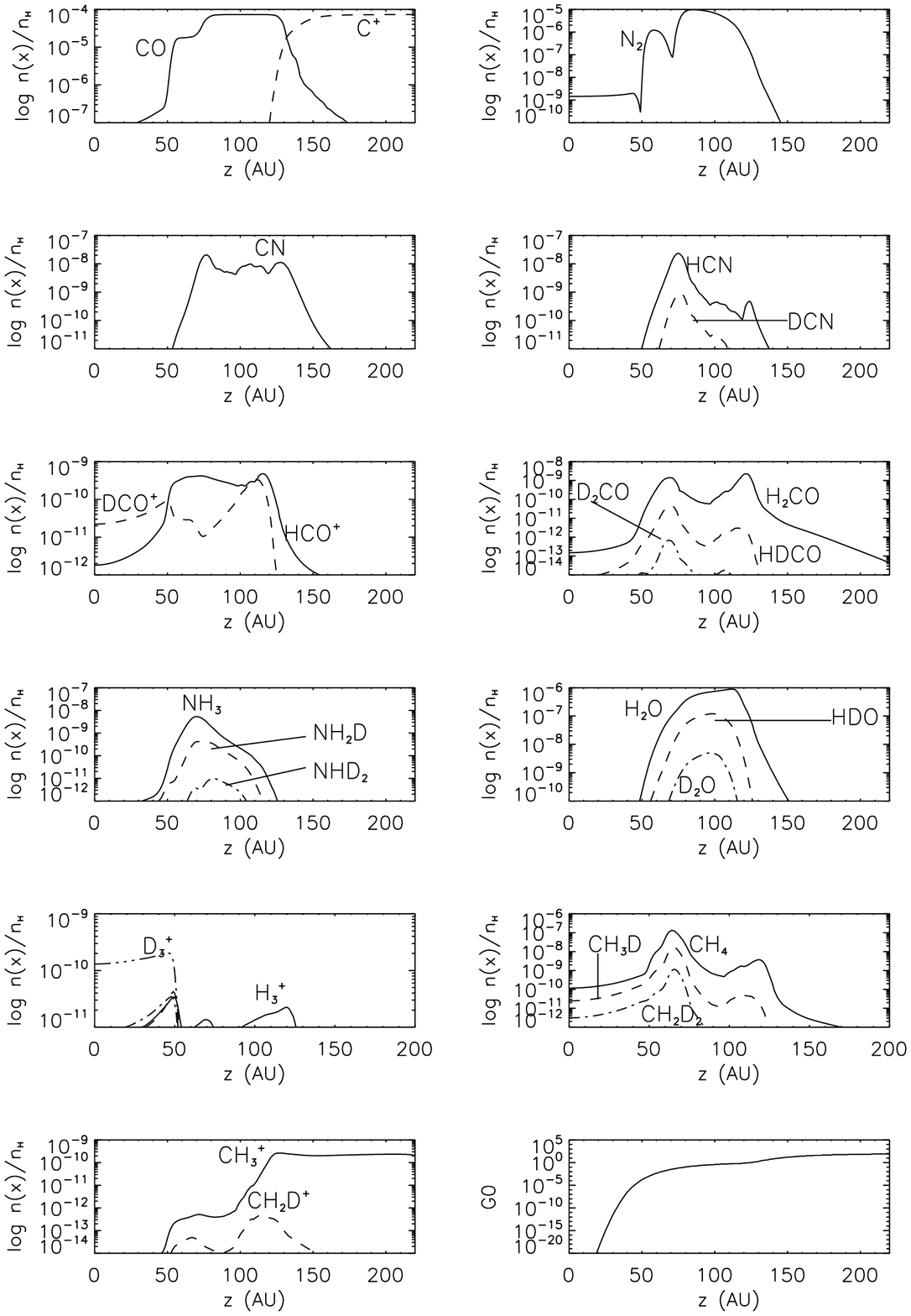}
\end{figure}

Figure~\ref{fig:frac_250_c} shows the fractional abundances
at 250 AU as calculated in Model C which includes both photodesorption
and CRH.  (Model D, which includes only photodesorption 
has a similar abundance distribution to Model C for $z$ $>$ 60 AU,
but with much lower gas phase abundances in the midplane.)
Photodesorption provides an 
efficient means of returning accreted species
to the gas, especially important for those molecules which are too strongly
bound to be affected by thermal desorption.  It produces
higher fractional abundances than Model B in the molecular
layer, where the UV field is high enough that 
molecules can be removed from the grains
but low enough that they can still 
survive in the gas.  Photodesorption also increases the
depth of the molecular layer.  

At $R$ = 250 AU,
photodesorption affects the vertical distribution 
of most molecules, even
those that can thermally desorb.  Without photodesorption, elements are 
gradually  removed from the gas by being incorporated into less volatile
molecules e.g.\ carbon is removed from CO and converted into hydrocarbon
ices,
and oxygen and nitrogen form H$_2$O and NH$_3$ ices respectively.  
When photodesorption
is included we see
large increases in the extent of the CO and N$_2$ layers.  
The most dramatic differences are for the less
volatile species.  For example, the peak fractional 
abundance of HCN increases to 2.5 $\times$ 10$^{-8}$ in Model C from 
6 $\times$ 10$^{-11}$ in Model B.  
NH$_3$ and H$_2$O show dramatic increases in both peak fractional
abundance and column density. All of these molecules can form
rapidly on the grain surface, where deuteration is also
efficient and so high column densities of HDO, D$_2$O, NH$_2$D etc.\
are also seen.  Similarly, N(H$_2$CO) is increased by grain formation
and desorption, although to a lesser extent than some of the other molecules.



\subsubsection{The effectiveness of cosmic ray heating}

Desorption due to cosmic ray heating (CRH) was first suggested as a means of
preventing the complete depletion of CO from the gas in cold, prestellar cores
\citep{ljo85}.  It was extended to cover other molecules in
molecular cloud models \citep{hh93}
and has been included in disk models by \citet{wl00} and \citet{wlab06}.  
Here we examine whether
its inclusion affects the calculated abundances and column densities.  
 
A comparison of the column densities calculated at $R$ = 250 AU
for Models A and B (Table~\ref{tab:cd})
shows that there is little difference for non-deuterated molecules.
However, several deuterated molecules are affected by CRH, because
this allows molecules to be exist in the gas in the cold midplane
of the disk where deuteration is efficient. 
For example, 
in Model B we see high abundances of DCO$^+$ at $z$ $<$ 50 AU 
(Figure~\ref{fig:frac_250_a}).  In Model A, CO is not present in the gas
in the midplane, and
therefore neither is DCO$^+$.  At $z$ $>$ 50 AU, where the
temperature is high enough for CO to thermally desorb, the temperature is
also high enough that deuteration is less efficient and
so little DCO$^+$ is present in Model A.  A similar effect is
seen for deuterated nitrogen-bearing molecules such as N$_2$D$^+$.
This difference in D/H ratios provides a possible means of
using the observations to differentiate between the models as
shown below.


\subsubsection{Vertical abundance distribution of the ices}

\begin{figure}
\caption{\label{fig:mantle_a}The fractional abundances as a function of
height, $z$ above the midplane for Models B and C.  
In Model B, the sharp decrease in 
abundance of CO and CH$_4$ at $z$ $\sim$ 60 - 70 AU, and
of H$_2$CO and isotopomers at $z$ $\sim$ 120 AU is due to the
onset of thermal desorption.  The carbon
atoms produced in the gas phase when CO is released and then
broken up are reaccreted on to the grains between 
60 and 140 AU where they form hydrocarbon ices. 
At $z$ $>$ 140 AU,  most accreted carbon atoms
are desorbed before they can react and hence the
abundance of hydrocarbons in the ice falls off.
In Model C photodesorption 
ensures that all molecules are removed by from the
grains by $z$ = 80 -- 100 AU. }
\epsscale{1}
\plottwo{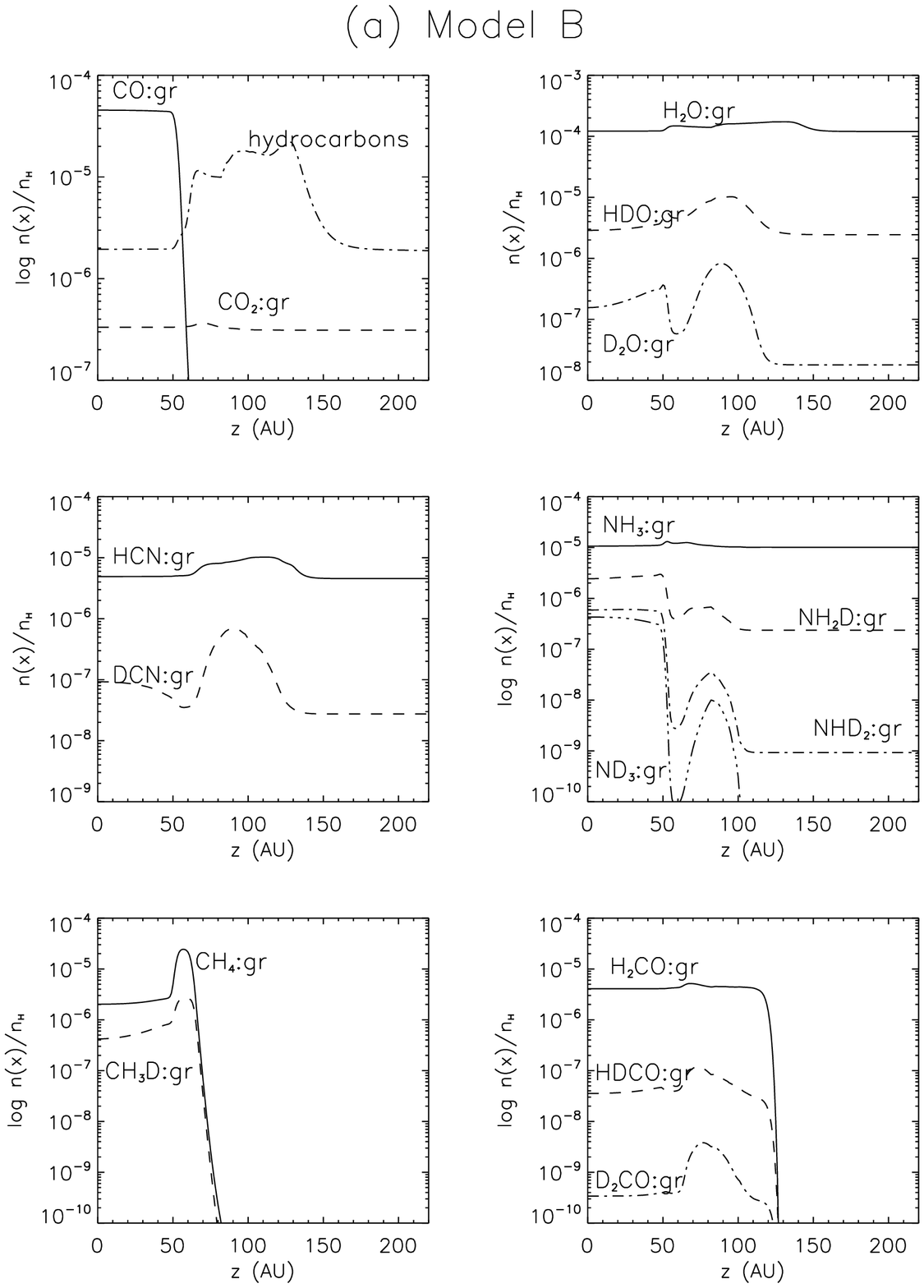}{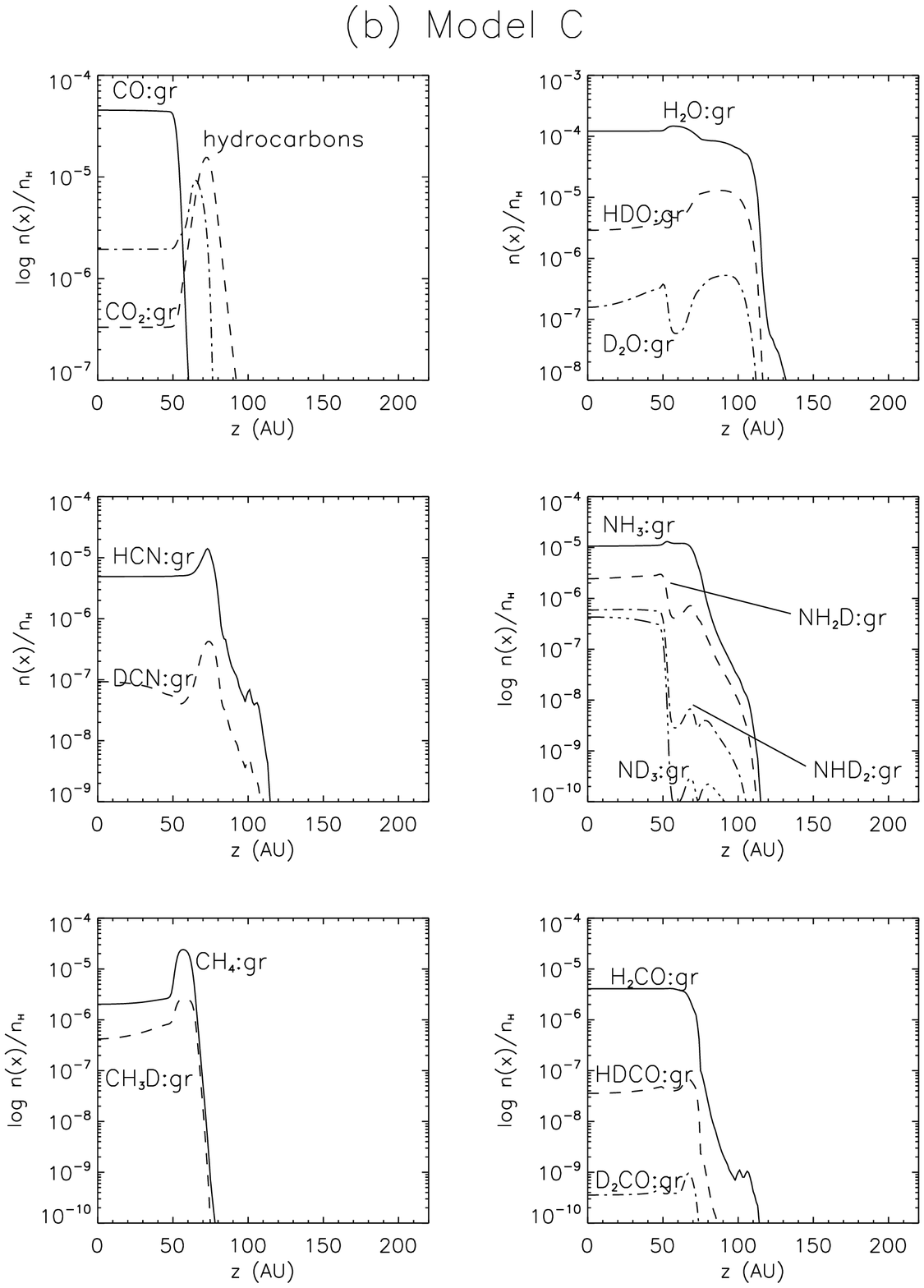}
\end{figure}

Figure~\ref{fig:mantle_a} shows how the mantle abundances vary with $z$ at
$R$ = 250 AU for Models B and C.  In both models, the most volatile species
have sharp decreases in abundance at $z$ $\sim$ 60 - 70 AU
where thermal desorption is able to return them to the gas.
This corresponds to the peak in gas phase abundance
seen in Figure~\ref{fig:frac_250_a}.  In Model B,
H$_2$CO and its isotopomers, which have a higher
binding energy than CO, N$_2$ and CH$_4$, are
removed from the grains at $z$ = 128 AU.  
In Model C, all species leave the grain between $z$ = 80 -- 100 AU
due to photodesorption.

In Model B many of the less
volatile ice species show little change in abundance with
height above the midplane, but there are some interesting
variations, especially for molecules that contain two or
more deuterium atoms.  In the case of D$_2$O, this molecule
forms on grains by the addition of deuterium atoms to an
accreted oxygen atom, but hydrogenation competes with 
the deuteration resulting in the production of H$_2$O and HDO
as well as D$_2$O.  The proportion of oxygen atoms ending up
in D$_2$O depends on the relative abundance of gaseous atomic 
hydrogen and deuterium.
At $z$ $<$  D/H is high, ensuring
that deuterium addition is efficient.  At higher $z$, photodissociation
begins to break down the main molecular reservoirs of hydrogen
and deuterium, namely H$_2$, HD and D$_2$.  Since there is
more hydrogen tied up in these molecules than deuterium, their
photodissociation results in a decrease in the atomic D/H
ratio, and hence deuterium addition to accreted oxygen atoms
becomes less competitive with hydrogen addition.  The rate
of formation of D$_2$O falls and we see this as a decrease in
its abundance above $z$ = 90 AU.  The same processes affect the
abundances of NHD$_2$ and ND$_3$ which are also formed on the grains.

Several molecules e.g.\ H$_2$O, HDO, HCN and DCN
show peaks in abundance between $z$ = 50 and 150 AU.
In this region, CO is abundant in the gas phase, where
it can be broken up into C$^+$ and oxygen atoms by reaction
with He$^+$.  Some of the oxygen atoms freezeout and 
react with hydrogen or deuterium, producing abundance
peaks in H$_2$O and HDO.  Similarly, the freezeout of C$^+$
can result in its reaction with nitrogen atoms on the grain surface
to form CN and subsequently HCN and DCN.  There is also a
peak in the abundance of hydrocarbons in the ice in this
region.  These form efficiently on the grains by the
reaction of carbon (produced by the photodissociation of
CO in the gas) and hydrogen.  Their abundance falls 
at $z$ $>$ 140 AU, where the majority of carbon exists
as gas phase ions.

Model C shows similar behavior at $z$ $<$ 80 AU.  The peak in 
CO$_2$ ice abundance just below this is due to its efficient formation
in the gas by the reaction of OH and CO.
The CO$_2$ then freezes out.  Photodesorption is active at this $z$ but
not sufficiently so to remove all of the accreted mantles.  Above
$z$ = 80 AU photodesorption dominates and the mantles are removed
by $z$ = 130 AU.

\subsubsection{\label{sec:h_be}How assumptions about the binding energy of 
hydrogen atoms affect the results}

As discussed in Section~\ref{sec:h_be} we have elected to use
a relatively high value of $E_D$(H) = 600 K \citep{ct02}.
This 
is almost double our previously assumed value of 350 K, and results
in more efficient H$_2$, HD and D$_2$ formation at temperatures
above 15 K, with correspondingly lower gas phase abundances of atomic
hydrogen and deuterium.  The binding energy not
only controls the residence time of hydrogen atoms on the grain,
but also affects their ability to react.  Hence the higher
binding energy used here will increase the time that 
a hydrogen atom remains on the surface but will also reduce the rate at which
hydrogenation processes can occur.  

To determine how $E_D$(H) affects the chemistry
we reran Model B at $R$ = 250 AU but this time with $E_D$(H) = 350 K
and $E_D$(D) = 371 K.  All other binding energies
and assumptions remain the same.  
The reduction in $E_D(H)$ increases the abundance of atomic
hydrogen at $z$ $>$ 50 AU.  For example, at $z$ = 70 AU (within 
the molecular layer) $x$(H) increases from 2.5 $\times$ 10$^{-7}$
for $E_D$(H) = 600 K to 4.7 $\times$ 10$^{-4}$ for $E_D$(H) = 350 K.
A similar
increase is seen in the atomic deuterium abundance.  The
column density of H doesn't change but that of D increases
from 6.5 $\times$ 10$^{15}$ cm$^{-2}$ to 3.5 $\times$ 10$^{16}$ cm$^{-2}$.

The changes in H and D abundance lead to changes in the abundances
and column densities of other gaseous molecules.  (The column
densities of molecules in the ice mantles do not show much
variation with the change in $E_D$(H)).  In the gas phase, with $E_D$(H) =350 K
N(H$_2$O) increases by $\sim$ a factor of 10, as does N(H$_2$CO), 
N(HDCO) and N(NH$_2$).  The largest increase is for CH$_3$OH
whose column density increases from 6.5 $\times$ 10$^5$ cm$^{-2}$ to 
2.2 $\times$ 10$^7$ cm$^{-2}$.  The reduction in binding energy
also changes some deuteration levels, most noticeably DCO$^+$/HCO$^+$
increases to 3.4 (up from 0.9 in Model B).

Using the lower value of $E_D$(H) in Model D changes the 
column densities even more.  There are large reductions in 
column densities of factors between 5 and 10 for many
of the deuterated molecules, including D$_2$O, 
all of the deuterated isotopomers of CH$_4$ and NH$_3$, DCN, 
and C$_2$D.  For example, N(D$_2$O) is reduced from 7 $\times$ 10$^{12}$ cm$^{-2}$
to 8.7 $\times$ 10$^{11}$ and N(DCN) from 1.2 $\times$ 10$^{12}$ cm$^{-2}$ to 
1.2 $\times$ 10$^{11}$ cm$^{-2}$.  DCO$^+$/HCO$^+$ is also
a factor of 4 higher than in Model D.

The choice of H-atom binding energy can therefore have
quite an effect on the calculated abundances.  We have
chosen the larger value since the results for a
few species are in better agreement with the observations:
in particular using $E_D$(H) = 350 K produces
DCO$^+$/HCO$^+$ ratios which are much higher than observed.



\subsection{The radial distribution of molecules}

\begin{figure}
\caption{\label{fig:rad_cd_a}The radial column density distributions calculated in Model B.}
\epsscale{0.8}
\plotone{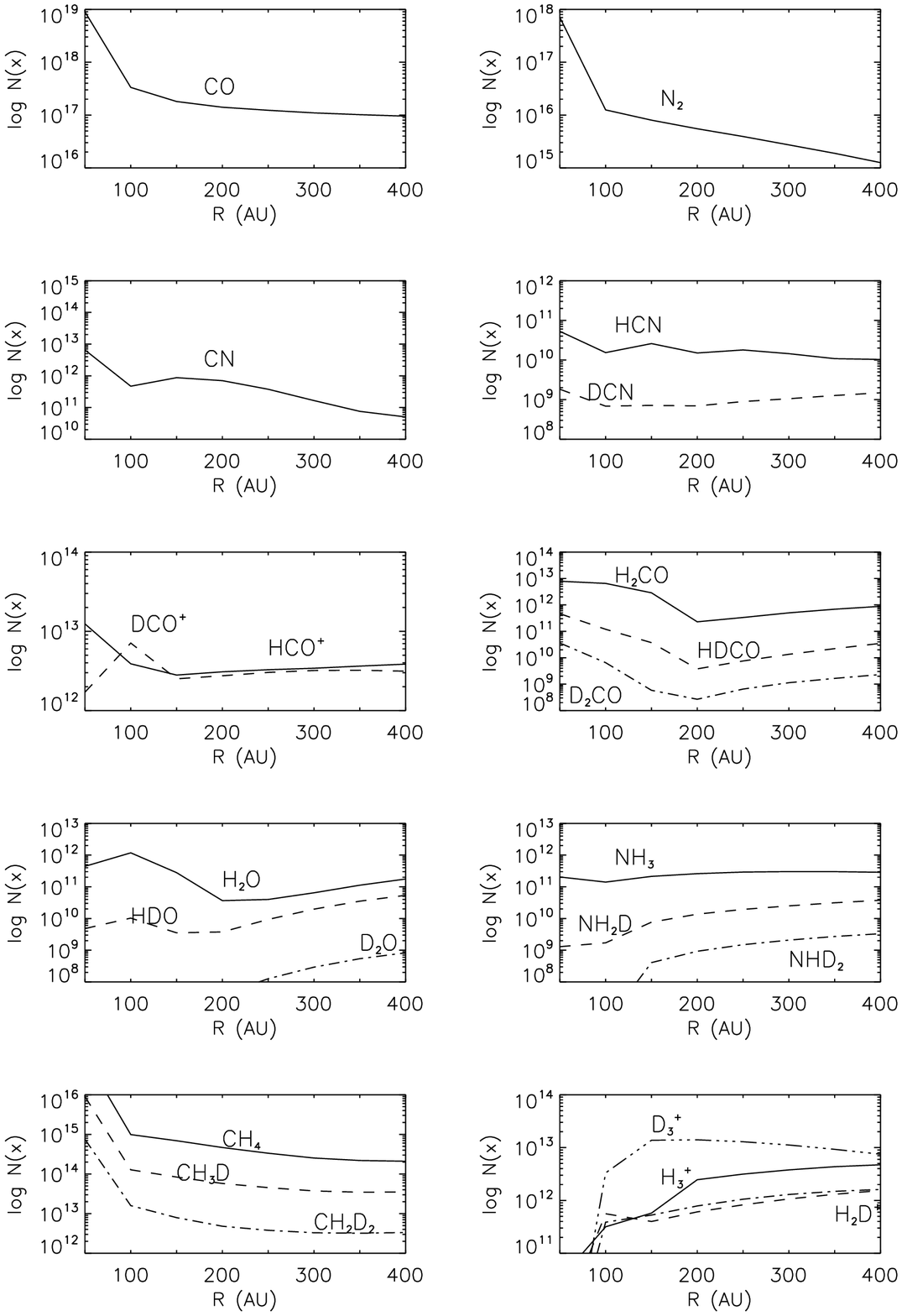}
\end{figure}

The radial distribution of molecular column densities, as calculated
in Model B, are shown in Figure~\ref{fig:rad_cd_a}.  N(CO) is fairly
flat with radius for $R$ $>$ 100 AU, but shows
a sharp rise at $R$ = 50 AU, where thermal desorption
of this molecule is efficient throughout the vertical extent of the disk.
N$_2$ has a more gradual increase because
of its slightly lower binding energy which means that it can desorb
at larger radii.    As discussed in the previous section, the
presence of these volatile species in the gas can lead to the production
of other molecules, which are not themselves thermally desorbed. 
For example, photodissociation of N$_2$ and CO can release
the elements required to form molecules such as HCN and HCO$^+$,
and the column densities of these species increase towards
the central star.


N(D$_3^+$) falls off with decreasing radius, because of the
increase in temperature which inhibits its formation from H$_3^+$.
The relative 
levels of deuteration in some other molecules such as HCO$^+$,
H$_2$O, HCN and NH$_3$ also decrease with decreasing $R$.
CH$_4$ and H$_2$CO do not show this decrease in deuteration,
because they can form from CH$_2$D$^+$, which is
less temperature sensitive due to the higher energy
barrier for the conversion of CH$_2$D$^+$ back into CH$_3^+$
($E_A$ = 370 K compared to $E_A$ = 230 K for the conversion of H$_2$D$^+$
into H$_3^+$).
Hence deuteration via reactions of CH$_2$D$^+$  is more efficient
in the warmer regions of the disk, than deuteration via H$_2$D$^+$.


The column density of H$_2$O falls off as the radius decreases until 
$R$ = 200 AU 
where it begins to rise again.  The decrease at $R$ = 50 AU
is due to the high photon field.

\begin{figure}
\caption{\label{fig:rad_cd_c}The radial column density distributions calculated in Model C.}
\epsscale{0.8}
\plotone{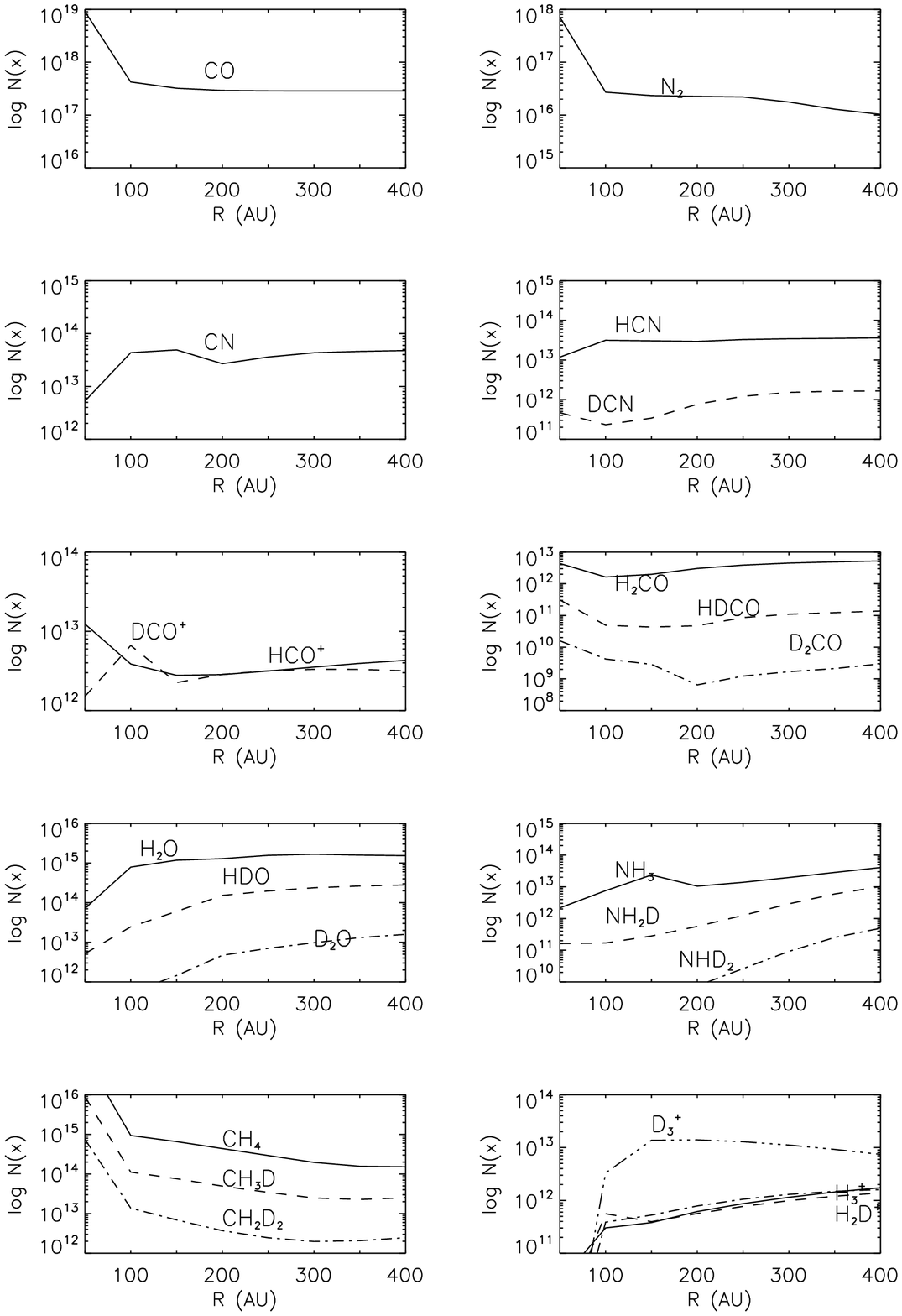}
\end{figure}

The radial distribution of column densities is also affected by photodesorption
(Model C: Figure~\ref{fig:rad_cd_c}), which increases the column densities of 
molecules at larger radii. 
Little difference is seen in the column densities
of volatile species in the two models at $R$ = 50 AU,
but at
larger radii photodesorption increases both N(CO) and N(N$_2$).
In contrast, some molecules are little affected at large radii by the 
inclusion of photodesorption.
For example, DCO$^+$ and HCO$^+$ have the same column densities in both
models, as do N$_2$H$^+$,  CH$_4$ and their isotopomers.

The distributions of HCN, DCN and CN are all related, with
CN being a photodissociation product of the other two molecules.
Their column densities in Model C
are fairly constant with radius for $R$ $\geq$ 200 AU,
but all three molecules show decreases in column density at smaller
radii, caused by the increase of photodissociation
rates closer to the star.  
The column densities in Model C are much higher than in Model B
and the radial distributions are different.
In Model B, CN, HCN, and DCN all peak towards the center of the disk,
whereas in Model C CN and HCN show a slight decrease.

\subsection{The effect of disk chemistry on D/H ratios}

\subsubsection{D/H ratios in the gas phase}
\citet{ah99a} found that chemical processing in the
disk can affect the molecular D/H ratios  and they
therefore suggested that the ratios observed in comets
do not just reflect the values set in the parent molecular
cloud, but are instead a combination of processing in the
cloud, during the infall phase and in the disk.  Similarly, we
find that gas phase D/H ratios are greatly changed
by the disk chemistry.  The situation for the grain mantles is 
a little more complicated, with disk chemistry having large
effects on some molecules, and very little on others.
This will be discussed below.


Table~\ref{tab:dh} lists the calculated D/H ratios at $R$ = 250 AU for each model
and compares them to the input values as determined in the molecular cloud model.
In the midplane very few molecules remain in the gas.  Those that
are present show a considerable increase in deuteration ratios compared to the input
values.  The effect is most marked for
multiply deuterated molecules.  Thus D$_3^+$/H$_3^+$ increases from 3.1 
$\times$ 
10$^{-3}$ in the molecular cloud
to 25 in the midplane of Models B and C, due to the high molecular depletions
in the disk.  This compares to H$_2$D$^+$/H$_3^+$ in the same models, 
which increases from an input value of 0.16 to $\sim$ 0.95.  The increase in the 
midplane
D$_3^+$/H$_3^+$ ratio is even higher in Models A and D where the
depletion is also higher due to the absence of CRH.

\begin{figure}
\caption{\label{fig:comp_dh}The D/H ratios calculated from the ratio of the
column density of the deuterated species to the column density of the non--deuterated
species, as a function of $R$.  Model B is shown in the solid line, with Model C as
the dashed line.}
\epsscale{0.95}
\plotone{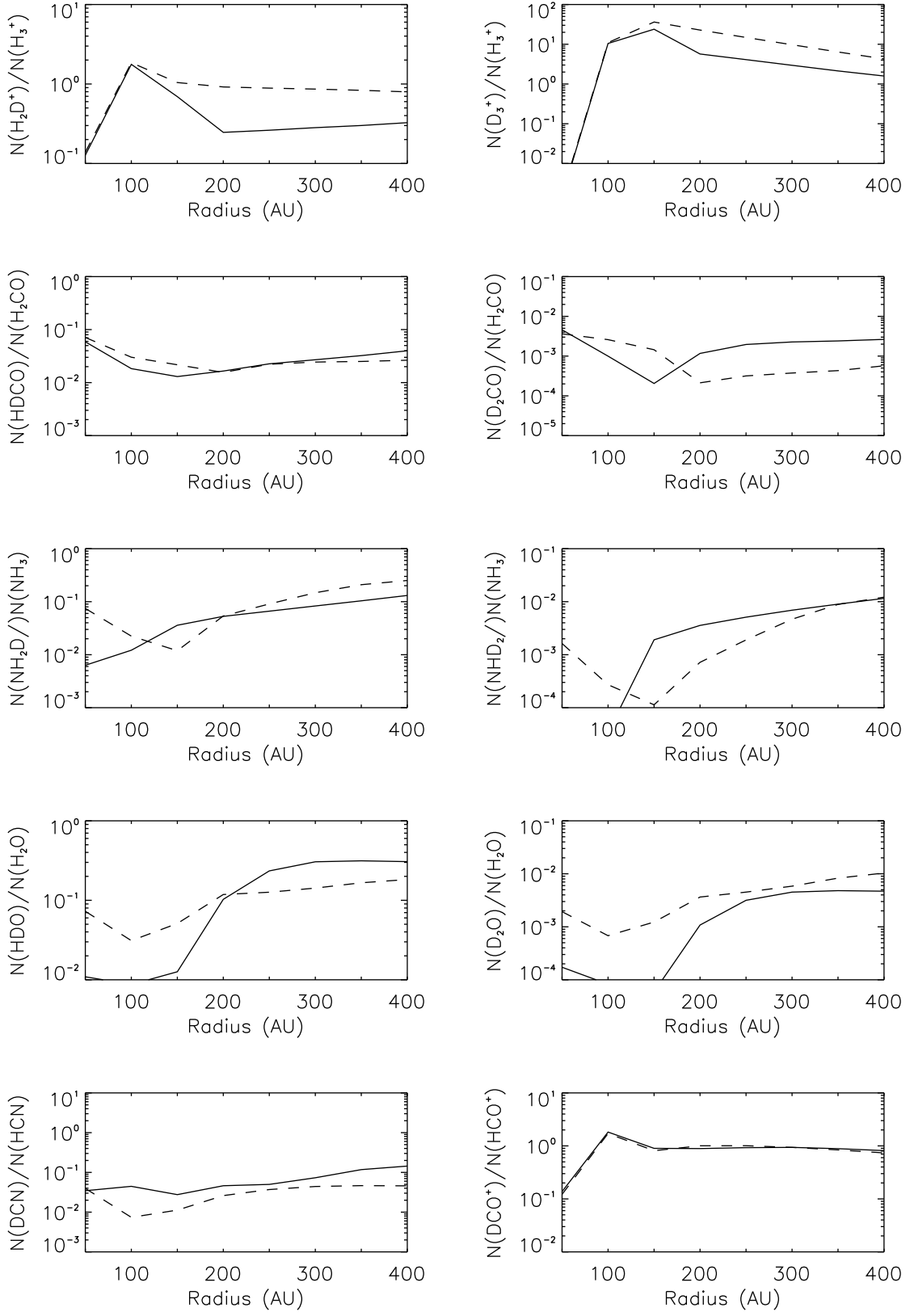}
\end{figure}

The radial variation of the ratios of the column densities
of deuterated to non--deuterated molecules in Model B 
(Figure~\ref{fig:comp_dh}) shows
a similar distribution to those of \citet{ah01}.  There is a
decrease in D/H with decreasing radius
for molecules such as NH$_3$, H$_2$O and HCO$^+$ 
whose deuteration depends on the isotopomers of H$_3^+$.  The
deuteration of H$_3^+$ decreases towards the star
because of the increase in temperature.  Molecules such as
H$_2$CO and CH$_4$ whose deuteration depends on CH$_3^+$
are less affected by the temperature increase, since the
barrier to the reverse reaction of
\begin{equation}
\hbox{CH$_3^+$} + \hbox{HD} \rightleftharpoons \hbox{CH$_2$D$^+$} + \hbox{H}_2
\end{equation}
at 370 K 
is higher than that for H$_2$D$^+$ + H$_2$ $\rightleftharpoons$ H$_3^+$ + HD
(220 K).
The ratio CH$_2$D$^+$/CH$_3^+$ decreases towards the center of the disk,
because there is a fall in HD in the molecular layer, caused partly by
the increase in photodissociation as the UV field increases and partly
by the efficient incorporation of deuterium into water and ammonia ices.

With the inclusion of photodesorption the picture changes.  This
mechanism can alter the molecular D/H ratios by injecting molecules formed
on the grains into the gas and therefore the gaseous D/H ratios partly depend
on the ratios in the ice mantles.
Some molecules do not show much
variation in the D/H ratio with photodesorption e.g.\ HDCO/H$_2$CO and
DCO$^+$/HCO$^+$ are roughly the same in both Models B and C.

In Models C and D the high abundance of molecules
in the molecular layer keep the abundances of H$_3^+$ and D$_3^+$
in this region low.  Hence most of the H$_3^+$ and D$_3^+$ are
in the midplane where the temperatures are low and the deuteration
is high leading to high D$_3^+$/H$_3^+$ ratios for these models. 
In Models A and B the higher depletion at $z$ $>$ 50 AU means that the
destruction rate of H$_3^+$ is lower and we see a secondary peak
in abundance for this molecule at $z$ $\sim$ 70 - 120 AU 
(Figure~\ref{fig:frac_250_a}).  Consequently N(H$_3^+$) is higher
and N(D$_3^+$)/N(H$_3^+$) is lower in these models.


In Model B we find that DCN/HCN decreases slightly with decreasing radius in 
agreement with \citet{ah01}, although our ratios are slightly higher
than theirs.
At large radii DCN forms from DCNH$^+$ which is produced by the reaction
of either DCO$^+$ or D$_3^+$ with HNC.  Hence DCN/HCN is
dependent on the deuteration of H$_3^+$. At $R$ = 50 AU,
neutral--neutral reactions become important with DCN forming
from \mbox{D + HNC $\longrightarrow$ DCN + H} and therefore the
DCN/HCN ratio is dependent on the atomic D/H ratio.  Another
important route to both DCN and HCN at small radii 
is the reaction of N with either CH$_2$D or CH$_3$.
DCN/HCN is also relatively constant with $R$ for Model C, 
where both molecules form on the grain surfaces and then
photodesorb.  In this case their ratio is therefore dependent on 
the gas phase atomic D/H.

In summary, chemical processing in the disk can affect the D/H
ratios of some molecules.  The effects is strongest for multiply
deuterated molecules, and for ammonia.  The radial variation
of molecular deuteration depends on the formation route
of the deuterated molecule, with species that are deuterated
via reactions with H$_2$D$^+$, HD$_2^+$ and D$_3^+$ showing
a decrease in the deuteration with decreasing radius, in contrast
to molecules deuterated by CH$_2$D$^+$, CHD$_2^+$ and CD$_3^+$.  If
photodesorption is included then the formation of deuterated
molecules on the grains becomes more important.

\subsubsection{D/H ratios in the ice mantles} 

In general, chemical processing in the disk has
less effect on the D/H ratios of the ices than
on the gas phase molecules.
Looking first at the midplane (Table~\ref{tab:dh}) 
we find
that the ratios calculated for the ices in Model A are close 
to the input values,
with the exception of DCN$:gr$/HCN$:gr$.  Model A
also shows no change in the fractional abundances of most 
ice molecules.
At the start of the model, most molecules have much higher
abundances on the grains than in the gas, and so the
freezeout of the gaseous molecules does not greatly affect
the grain mantle abundances.  The exceptions are for CO, N$_2$,
HCN and DCN.  For CO and N$_2$ the initial gas phase abundances
are high.  In the case of HCN and DCN these molecules form on the
grains from the freezeout and hydrogenation/deuteration of CN.

CRH causes some changes in the ice deuteration in the midplane
in Models B and C.  As in the gas we find that the
biggest changes are for multiply deuterated molecules. 
The break up of desorbed CO and N$_2$ provides
carbon ions and oxygen and nitrogen atoms which
can freezeout and react on the grain surface, most
likely with hydrogen or deuterium.  The
relative efficiency of these two processes depends on the
atomic D/H ratio, which in the midplane is very high.  Hence
multiply deuterated ices are easily formed in this region.
Ammonia ice is particularly highly deuterated because a lot
of it is produced in the disk (due to the high input
abundances of both gaseous and solid N$_2$  which provide
a plentiful source of nitrogen atoms), whereas for H$_2$O and
CH$_4$ the majority of these ices are produced in the molecular
cloud phase and the disk chemistry makes a much lower
contribution to both their abundances and their deuteration.



If we calculate the molecular D/H ratios from the
column densities of the ices, we again find that Models A and D
show the least change from the molecular cloud values.
In Models B and C the
column density ratios show similar behavior to the
midplane ratios, with multiply deuterated species 
and the isotopomers of ammonia being
most affected.


In summary we find that the inclusion of a non--thermal
desorption process, such as CRH, which is effective
in the midplane will affect the D/H ratios of some
of the ice molecules and that this effect increases
for multiply deuterated molecules.  Such a process would
be expected to alter the D/H ratios of the material that
is later incorporated into Solar System bodies such as comets. 
In the absence of
such a non--thermal desorption process, the ice mantles retain
the D/H ratios set in the molecular cloud and are unaffected
by the disk chemistry.  The exception to this is
ammonia, which is predicted to be highly deuterated in the ice
by all of our models.  At smaller radii, where thermal desorption
is efficient, we would expect that disk chemistry will
play a more important role in determining the D/H ratios
of the ices.  

There are limitations to our model.  We have not incorporated
chemical processing of the grain mantles by cosmic rays or UV photons.
  UV will not be important in the
heavily shielded midplane and so can be safely ignored in this
region, but cosmic rays can penetrate and may have a chemical effect.
Whether or not this can affect the D/H ratios is unclear.  A further
complication ignored here is processing of the grain mantles during
the collapse phase as the molecular cloud material is accreted into the
disk.  An accretion shock forms at the surface of the disk and it is possible
that the grains will lose their mantles as they travel through this.
\citet{lunine91, cc97} find that grains will lose their water
ice coatings if they are infalling at radii of less than 30 AU, with the
effect decreasing with increasing radius \citep{lunine91}.  The grains
will regain their ice mantles once they have passed through the shock,
but processing in the shock could alter the composition and deuteration
levels of the ices.
Both these papers suggest that at least some of the material in comets
comes directly from the interstellar medium.  
At the radii under consideration here, it seems likely based on 
\citet{lunine91,cc97} that at least some of the grains will retain their
volatiles as they are accreted into the disk.  
At present it is unclear what effect
the accretion shock would have on the D/H ratios of the ice mantles
on grains at smaller radii, where these mantles may have been desorbed
in the shock
and the released molecules processed in the gas before re--accreting 
onto the grains.

It has been suggested that comets are composed of interstellar
grains that have undergone relatively little processing in the
Solar Nebula \cite[see review by][and references therein]{irvine00}.  One reason
for supposing this is the similarity between the chemistry of
comets and interstellar clouds, and in particular the similarity
between the molecular D/H ratios.  
Dynamical studies of comets have shown that short period comets
formed in the trans--Neptunian region \citep{duncan88} whereas
long period comets formed closer to the Sun and then had
their orbits perturbed by interaction with the giant planets, to
eject them outwards to the Oort cloud.  Our model is of the outer
disk, with an inner radius well outside the region where
long period comets - the only ones for which we have observed D/H ratios - 
are likely to have formed.  Still we include the comet data here
for comparison, while recognizing that models of the inner disk are
required to properly cover the comet formation region.
In Hale--Bopp, DCN/HCN = 2.3 $\times$ 10$^{-3}$ \citep{meier98a}.
HDO/H$_2$O = 3.3 $\times$ 10$^{-4}$  in Hale--Bopp \citep{meier98a}
and 2.9 $\times$ 10$^{-4}$ in Hyakutake \citep{bm98}.  Our model results
are close to the input values with
DCN$:gr$/HCN$:gr$ $\sim$ 6.6 $\times$ 
10$^{-3}$ and HDO$:gr$/H$_2$O$:gr$ $\sim$ 2.7 $\times$ 
10$^{-2}$ in the midplane
at $R$ = 50 AU (Model D).   
The DCN$:gr$/HCN$:gr$ ratio is consistent with the
observations but HDO$:gr$/H$_2$O$:gr$ 
is much higher than the observed
value.  The discrepancy could be a result of too efficient deuteration on
grain surfaces, or it could be because our model considers
a region further out in the disk than the comet formation radius.
Processing at a small radii (with higher temperatures) could reduce the 
calculated HDO/H$_2$O value.

\section{\label{sec:obs}Comparison to observations}

One reason for modeling the outer parts of the protoplanetary disk
is that most current observations in the millimeter and submillimeter
probe this region and therefore can provide a test of the models.
Three disks in which several molecules (including deuterated forms)
have been observed are LkCa15, DM Tau and TW Hya.
\begin{itemize}
\item {\it TW Hya:} This object is the closest classical T Tauri star known. Its disk
is almost face on with a radius of 200 AU. It
is relatively old with an estimated age of 5 - 20 Myrs.  It has
a mass accretion rate, $\dot M$ = 10$^{-9}$ -- 10$^{-8}$ M$_\odot$ yr$^{-1}$
\citep{kastner02}.  The disk mass is estimated from
continuum observations to be 3 $\times$ 10$^{-2}$ M$_\odot$ \citep{wilner00}.

\item {\it LkCa15:}
This source is one of the strongest T Tauri emitters in the millimeter found in the
survey of \citet{beckwith90}.  It is found located in  Taurus and has an age of $\sim$ 3--10 Myrs
\citep{simon00,thi01}.
The disk mass is $\sim$ 0.03 M$_\odot$ and the mass accretion rate
is 10$^{-9}$ M$_\odot$ yr$^{-1}$ \citep{hartmann98}.  The disk radius is 650 AU
\citep{simon00}.  The stellar mass is 1 M$_\odot$
\citep{simon00} and its temperature is 4365 K \citep{muz00}

\item {\it DM Tau:}  This object is also found in Taurus.  It has an age of $\sim$ 5Myrs.
Its stellar mass is 0.55 $M_\odot$ \citep{simon00} and its stellar temperature
is 3720 K \citep{gd98}.  The disk has a mass of 0.03 M$_\odot$ \citep{gd98},
a radius of $\sim$ 800 AU \citep{simon00} and a mass accretion rate
of 10$^{-8}$ M$_\odot$ yr$^{-1}$ \citep{hartmann98}.

\end{itemize}

We compare our results with the available observations of
both deuterated and non--deuterated species.  
The column densities derived from observations are very sensitive to 
assumptions made about the source.  Single disk observations cannot 
resolve the disks and assumptions about the density and temperature
structure of the disks must be made in order to determine abundances
and column densities.  For DM Tau, \citet{dutrey97} determined the gas
density distribution of a geometrically thin disk in hydrostatic
equilibrium and used this to derive the average fractional abundances
with respect to H$_2$, assuming that the fractional abundances were
the same everywhere in the disk.  \citet{aikawa02} used the model and
data from \citet{dutrey97} and integrated vertically to determine the
column densities quoted in Table~\ref{tab:obs}.  For LkCa15 both
single dish and interferometric data is available.  \citet{qi01} used
the Owens Valley Radio Interferometer (OVRO) array to observe this source and
derived beam average column densities.  The resolution of the array
($\sim$ 300 AU at the distance of LkCa15) is such that the source is
just resolved.  \citet{thi04} present single dish observations of LkCa15.  
They derive the column densities assuming that the radius of the
disk is 450 AU.  The disk radius can have a large effect on the
calculated column densities with a change from $R_1$ to $R_2$ scaling
the column densities by $(R_1/R_2)^2$.  The single dish column
densities are significantly higher than those from the interferometer
data due to the differences in the assumptions used to determine these
numbers.

Several deuterated molecules are now claimed to have been
detected in protoplanetary disks.  However, it
should be noted that some of these detections have
recently been disputed.  \citet{guilloteau06} re--analyzed
the spectra of HDO and H$_2$D$^+$ in DM Tau \citep{cec05} and found only
a 2-$\sigma$ detection of H$_2$D$^+$ and no evidence of
HDO in this disk.  They attributed the discrepancy between
their work and that of the original authors to the use of an over--estimation
of the continuum flux and the wrong systematic velocity by the latter.  
In the following
discussion we have elected to use all the available data on deuterium
molecules in disks for comparison with our models, including that
which is still under dispute.

The observations referred to here are of the outer disk, where
the radius is greater than 50 -- 100 AU.  Here
we use our results at $R$ = 250 AU as representative of the outer disk
model and compare these to the observations.
The results for Models B, C and D, together with the observations are
given in Table~\ref{tab:obs}.  Good agreement is defined as less than
a factor of 5 difference between the models and the observations.
We have not included Model A in this table since the column densities
for this model are consistently lower than the others and much
lower than observed, indicating that non--thermal desorption is
required to account for the observations.

For each source we find that Models C and D give better agreement with the
observations than Model B.  In particular, 
strongly bound molecules such as HDO require
the presence of photodesorption in order to remain in the gas phase.

\citet{cec05} observed HDO in absorption in DM Tau using the
JCMT and determined N(HDO) = 1.6 $\times$ 10$^{13}$ cm$^{-2}$.
In Model B, N(HDO) is too low at 9.3 $\times$ 10$^9$ cm$^{-2}$
to account for the observations, but in Models C and D photodesorption
can keep a much higher abundance of HDO in the gas.  These
models have N(HDO) = 2.0 $\times$ 10$^{14}$ cm$^{-2}$, somewhat
higher than the observed value, but demonstrating that photodesorption
is an efficient means of returning HDO (and other molecules) to the gas.

\cite{dom05} also looked the effects of photodesorption on water
and its isotopomers.  Using a simplified model they calculated the
column density of H$_2$O averaged across a face on disk 
to be 1.6 $\times$ 10$^{15}$ cm$^{-2}$, 
in excellent agreement with our model.  
Combining their model results with 
the observations of \citet{cec05}, they estimate HDO/H$_2$O $\sim$ 0.01.
Our models calculate much higher ratios with HDO/H$_2$O = 0.13 in 
Models C and D.  These calculated ratios are also much higher
than are observed in Solar System objects and may point to
deuteration on the grains being less efficient than assumed here.

There is another reported detection of HDO in a protostellar disk.
\cite{qi01} observed HDO in LkCa15 using the OVRO interferometer and found
N(HDO) = 2 -- 7 \x 10$^{14}$ cm$^{-2}$, with an intensity peak that
was offset from the central star.  As in DM Tau, models that include photodesorption
give the best agreement with the observations.

\cite{qi01} also observed DCN in LkCa15 with N(DCN) $\sim$ 10$^{13}$ cm$^{-2}$
and DCN/HCN = 0.01.  Model B calculates 
a low value of N(DCN) = 8.9 \x 10$^{8}$ cm$^{-2}$, but Models C and D have
N(DCN) = 1.2 \x 10$^{12}$ cm$^{-2}$ (DCN/HCN = 0.036), in good
agreement with the observations.


H$_2$D$^+$ has been observed in DM Tau \citep{cec04}
with  a column density of 8.8 \x 10$^{12}$ cm$^{-2}$,
and a midplane fractional abundance
$x$(H$_2$D$^+$) = 3.4 \x 10$^{-10}$.
In all our models we find $x$(H$_2$D$^+$) $\sim$ few \x 10$^{-12}$ 
in the midplane for $R$ $\geq$ 100 AU, and N(H$_2$D$^+$)
$\sim$  few \x 10$^{11}$ cm$^{-2}$.  Our calculated N(H$_2$D$^+$)
is therefore a factor of 10 lower than
observed and our midplane fractional abundance more than
a factor of 100 lower.

DCO$^+$ and HCO$^+$ have both been detected in two sources:
DM Tau \citep[DCO$^+$/HCO$^+$ = 4 \x 10$^{-3}$;][]{guilloteau06}
and TW Hya \citep[DCO$^+$/HCO$^+$ = 0.035;][]{vd03}.  
The formation of these molecular ions depends on the presence of CO
in the gas. The column density of CO in DM Tau is 5.7 \x 10$^{16}$ cm$^{-2}$
\citep{dutrey97} in good agreement with all our models.
\cite{guilloteau06} do not
give a column density for DCO$^+$, but in TW Hya N(DCO$^+$) = 3 \x 10$^{11}$
cm$^{-2}$ compared to the calculated values of $\sim$ 3.1 \x 10$^{12}$ in 
Models B and C and 4.8 \x 10$^{11}$ in Model D.  In the same source, N(HCO$^+$)
= 8.5 \x 10$^{12}$ cm$^{-2}$ and the model values, which range from
2.8 - 3.3 \x 10$^{12}$ cm$^{-2}$, are in good agreement with this.
In our models the desorption
process that affects the calculated ratios most is CRH.  This 
allows CO to be present in the cold midplane where deuteration is most
efficient.  When CRH is included N(DCO$^+$) is high and 
our calculated DCO$^+$/HCO$^+$ $\sim$ 1 (Models
B and C).  Without CRH (Model D) both N(DCO$^+$) 
and  the ratio of 0.18 are in agreement with the TW Hya
observations.  However for this model DCO$^+$/HCO$^+$ is 42 times higher 
than seen in DM Tau.  The observations
and modeling of the DCO$^+$/HCO$^+$ ratio suggests that CO is not present
in the midplane, but instead does not begin to become abundant
until the disk is warm enough that deuteration is not so efficient,
as happens in Model D.
Based on this we suggest that CRH is not acting in disks, or if
it is, it is less efficient than assumed in our (and other) models.

\citet{qi03} observed N(N$_2$H$^+$) = 3.1 $\times$ 10$^{13}$ cm$^{-2}$
in LkCa15.  The highest column density found in any of our models
is 2.8 $\times$ 10$^{10}$ (Model C), $\sim$ 1100 times lower
than observed. 
The discrepancy is even
larger for Model D, where N(N$_2$H$^+$) = 1.1 \x 10$^{10}$ cm$^{-2}$.
The increase in N(N$_2$H$^+$) in models that include CRH
suggests that a high column density of N$_2$H$^+$ requires that
N$_2$ is present in the midplane of the disk.

So from a comparison of observations with our models we can
see that we require N$_2$ to be present in the midplane, but CO to
be absent.  In the current models the
binding energies of these two molecules are similar and therefore
they have similar desorption rates.  In order to account for the observations
we would need $E_D$(N$_2$) to be
somewhat smaller than $E_D$(CO).  The
values we use are taken from \cite{oberg05} who measured
the binding energy of N$_2$ on N$_2$ ice, CO on CO ice
and the binding energies of both molecules on a mixture
of CO and N$_2$ ices.  However, in the interstellar medium
and protostellar disks the main component of the ice mantles
is water and it is possible that the binding energies of
N$_2$ and CO to water ice are very different from those
measured by \cite{oberg05}.
Various theoretical and experimental
work points towards this being the case.  For example, \cite{hc06}
take $E_D$(CO) and $E_D$(N$_2$) to be 1150 K and 1000 K respectively,
based on the laboratory work of \cite{collings03} and \cite{ayotte01}.
Calculations by \cite{alhalabi04} find $E_D$(CO) to be 
1010 K on amorphous ice and 1215 K on crystalline ice.
These values are rather more than the \citeauthor{oberg05}\ values.
Increasing the binding energies will reduce the efficiency of both
thermal desorption and CRH (for which rates are calculated
using $E_D$).  Also increasing the relative value of $E_D$(CO) compared
to $E_D$(N$_2$) will mean that N$_2$ could be desorbed in regions
where CO is retained on the grains.  For example, using $E_D$(CO)
= 1200 K and $E_D$(N$_2$) = 1000 K, we find midplane
abundances $x$(N$_2$) = 3.2 \x 10$^{-10}$ and $x$(CO) = 5.2 \x 10$^{-10}$
at $R$ = 250 AU, compared to 1.4 \x 10$^{-9}$ and 6.0 \x 10$^{-8}$
respectively using the \citeauthor{oberg05} binding energies.  
In terms of column densities, the higher binding energies 
have N(N$_2$H$^+$) = 6.2 \x 10$^{10}$ cm$^{-2}$ (an increase
of a factor of 2.4 over the value in Model C), while
N(CO) = 1.4 \x 10$^{17}$ cm$^{-2}$ and DCO$^+$/HCO$^+$ = 0.23.
Thus with the higher binding energies, N(CO) is still in 
agreement with the observations and N(N$_2$H$^+$) increases but is
still a factor of $\sim$ 500
lower than the observed value in LkCa15.  A higher column
density could be achieved if N$_2$ is more efficiently
desorbed in the midplane.

Even with the increased CO binding energy, DCO$^+$/HCO$^+$ is
still much higher than observed, because CRH can still
desorb some CO in the midplane.  
Another possibility is that CO does not desorb by CRH
as efficiently as we assume here. 
Our rates are based on the work of \citet{hh93},
but recent work by \cite{bj04} suggests that
theses rates maybe an over-estimate for CO and an underestimate
for H$_2$O.
According to Bringa and Johnson the desorption rate for CO is
a factor of 10 lower than estimated by \cite{hh93} which 
would reduce the abundance of CO in the midplane considerably.
There is no data for N$_2$, but if it is able to desorb at
the Hasegawa \& Herbst rate, then this would allow
there to be a midplane layer of N$_2$ that is also
lacking in CO.

On the basis of our modeling we find that efficient
non--thermal desorption such as photodesorption must
be active in the surface and molecular layers to account
for the observed column densities of molecules such as HDO
and HCN which would otherwise be expected to be accreted
onto the grains.  The observed DCO$^+$/HCO$^+$ ratio indicates that
CO is not present in the midplane where deuteration is efficient,
ruling out desorption of CO in this region that
is as efficient as the CRH included here.  However, some
desorption must be occurring in the midplane 
to account for observations of N$_2$H$^+$.  Overall Model D
gives the best agreement with observations, but in order
to account for N$_2$H$^+$, N$_2$ must be able to desorb in the
midplane.

\section{Summary}

We have presented results from new models of the
deuterium chemistry in disks which include grain mantle chemistry 
and multiply deuterated molecules, and have considered how
desorption processes affect the calculated abundances
and the molecular deuteration.  Our main findings are as follow:
\begin{enumerate}
\item Very high levels of deuteration can be achieved in the 
midplane, where freezeout removes most molecules from the gas.  
\item Efficient desorption is required in the molecular layer to 
account for the observations of molecules such as HDO.  
In common with other authors e.g.\ \cite{wl00} and \cite{dom05},
we find that photodesorption is able to account for the
observed column densities.
\item Chemical processing in the disk, including grain surface
chemistry, can make great
changes to the deuteration of gas phase molecules compared to the
input molecular cloud values.  This is partly a consequence of
processing in the gas and partly due to the formation of molecules
on the grains and their subsequent desorption.  We find that
DCO$^+$/HCO$^+$ is very sensitive to the desorption processes
included in the models.  From a comparison of observational data
for this molecule and
the models we find that CO cannot be present in the midplane,
suggesting either that the rate of cosmic ray desorption of
this molecule is less than assumed in these models, or that
its binding energy on water ice is much higher than the value
adopted here.
\item From comparison of the model results in the midplane
at 50 AU we find that some molecules found in comets have
D/H ratios which reflect the input molecular cloud abundances, 
but that others, notably NH$_3$ have undergone considerable
chemical processing in the disk.  Based on this it
seems that comets could be comprised of some molecules
that have their origin in the parent molecular cloud and others 
that were formed in the disk.
\end{enumerate}

The deuterium chemistry provides a means of testing the models
against observations and of distinguishing among the various
processes that might be acting in the disk.  As observations 
continue to be acquired and gain in sensitivity and resolution, 
the ability to use models and observations together to understand the
chemistry of protoplanetary disks will be greatly enhanced.

\acknowledgements
This research was conducted at the Jet Propulsion Laboratory,
California Institute of Technology under contract with the
National Aeronautics and Space Administration.  Partial
support was provided by the NASA TPF Foundation Science
Program.

\bibliography{/home/willacy/papers/disk_turb/disk}


\begin{deluxetable}{lrcclrc}
\tablecolumns{7}
\tablewidth{0pt}
\tablecaption{\label{tab:be}
The binding energies ($E_D$) used
to determine the thermal desorption rates of the abundant
mantle species. }
\tablehead{
\colhead{Species} & \colhead{$E_D$ (K)} & \colhead{Reference} &
\colhead{} & \colhead{Species} & \colhead{$E_D$ (K)} & \colhead{Reference}
}
\startdata
H        & 600  & 1 & & D        & 621  & 2 \\
H$_2$    & 315  & 8 & & C        & 800  & 3 \\
CH       & 645  & 9 & & CH$_2$   & 956  & 9 \\
CH$_3$   & 1158 & 9 & & CO       & 855  & 4 \\
CO$_2$   & 2860 & 5 & & H$_2$CO  & 1760 & 3 \\
CH$_3$OH & 4240 & 7 & & O        & 800  & 3 \\
O$_2$    & 1210 & 3 & & OH       & 1259 & 9 \\
H$_2$O   & 4820 & 6 & & N        & 800  & 3 \\
N$_2$    & 790  & 4 & & NH       & 604  & 9 \\
NH$_2$   & 856  & 9 & & NH$_3$   & 3080 & 5 \\
\enddata
\tablerefs{(1) \citet{ct02,ct04}, (2) \citet{caselli02}, (3) \citet{ta87},
(4) \citet{oberg05}, (5) \citet{sa90}, (6) \citet{sa88}, (7) \citet{sa93},
(8) \citet{rh00}, (9) \citet{ar77}}
\end{deluxetable}

\begin{deluxetable}{ll}
\tablecolumns{2}
\tablewidth{0pt}
\tablecaption{\label{tab:cloud}Elemental abundances used in the model.
Abundances are given with respect to the total abundance of hydrogen
atoms, $n_H$ = 2 $n$(H$_2$) + $n$(H).  Initially all deuterium is
contained in HD, and all hydrogen in H$_2$.}
\tablehead{
\colhead{Element} & \colhead{Abundance}\\
& \colhead{$n(x)/n_H$}
}
\startdata
H$_2$ & 0.5\\
He & 0.14 \\
HD & 1.6 \x\ 10$^{-5}$ \\
O & 1.76 \x\ 10$^{-4}$ \\
C$^+$ & 7.3 \x\ 10$^{-5}$ \\
N & 2.14 \x 10$^{-5}$ \\
Fe & 3.0 \x 10$^{-9}$ \\
Mg & 7.0 \x 10$^{-9}$\\
\enddata
\end{deluxetable}

\begin{deluxetable}{llll}
\tabletypesize\footnotesize
\tablecolumns{7}
\tablewidth{0pt}
\tablecaption{\label{tab:input}Input abundances for the disk model
as determined by a molecular cloud model at 1 Myrs.  
The abundances are given relative to the total abundance
of hydrogen $n_H$ = 2 $n($H$_2$) + $n$(H).  For comparison
we have included the abundances observed at the cyanopolyyne peak 
in the molecular cloud TMC--1.}

\tablehead{
\colhead{Molecule} & \multicolumn{2}{c}{Fractional abundance} & \colhead{TMC--1} \\
 & \colhead{Gas} & \colhead{Grain} & \colhead{Gas phase}\\
 &               &                 & \colhead{observations}
}
\startdata
H             & 3.8 (-5)  & \nodata   & \\
D             & 7.9 (-7)  & \nodata   & \\
HD            & 1.1 (-5)  & \nodata   & \\
H$_3^+$       & 3.5 (-9)  & \nodata   & \\
H$_2$D$^+$    & 5.7 (-10) & \nodata   & \\
HD$_2^+$      & 8.3 (-11) & \nodata   & \\
D$_3^+$       & 1.1 (-11) & \nodata   & \\
CO            & 4.2 (-5)  & 3.5 (-6)  & 8.0 (-5)$^1$ \\
CO$_2$        & 1.7 (-8)  & 3.1 (-7)  & \\
C$_3$O        & 1.0 (-10) & \nodata   & 1.0 (-10)$^1$ \\
HCO$^+$       & 2.9 (-9)  & \nodata   & 8.0 (-9)$^1$ \\
DCO$^+$       & 2.1 (-10) & \nodata   & 1.6 (-10)$^2$ \\
H$_2$CO       & 5.3 (-8)  & 4.0 (-6)  & 5.0 (-8)$^1$ \\
HDCO          & 2.2 (-9)  & 2.8 (-8)  & 3.0 (-9)$^3$ \\
D$_2$CO       & 5.4 (-11) & 2.7 (-10) & \\
CH$_3$OH      & 1.5 (-10) & 3.9 (-8)  & 3.0 (-9)$^{1,4}$ \\
CH$_3$OD      & 3.8 (-12) & 1.7 (-10) & 8.0 (-11)$^3$  \\
CH$_2$DOH     & 1.1 (-11) & 1.9 (-9)  & \\
O             & 3.3 (-6)  & \nodata   & \\
O$_2$         & 6.0 (-8)  & \nodata   & \\
OH            & 4.5 (-8)  & \nodata   & 2.0 (-7)$^1$ \\
OD            & 3.2 (-8)  & \nodata   & \\
H$_2$O        & 8.1 (-8)  & 1.2 (-4)  & \\
HDO           & 3.6 (-9)  & 2.4 (-6)  & \\
D$_2$O        & 1.6 (-11) & 1.8 (-8)  & \\
N             & 4.5 (-7)  & \nodata   & \\
N$_2$         & 2.2 (-6)  & 7.3 (-8)  & \\
NO            & 1.1 (-7)  & \nodata   & 3.0 (-8)$^1$ \\
CN            & 2.5 (-8)  & \nodata   & 5.0 (-9)$^4$ \\
HCN           & 1.5 (-8)  & 4.6 (-6)  & 2.0 (-8)$^1$ \\
DCN           & 4.9 (-10) & 2.8 (-8)  & 2.2 (-10)$^{2,3}$ \\
HNC           & 1.0 (-8)  & 7.2 (-7)  & 2.0 (-8)$^1$ \\
DNC           & 1.2 (-10) & 3.0 (-9)  & 3.0 (-10)$^{3,5}$ \\
C$_3$N        & 2.9 (-9)  & \nodata   & 6.0 (-10)$^9$ \\
HC$_3$N       & 5.8 (-8)  & 3.7 (-7)  & 2.0 (-8)$^4$ \\
DC$_3$N       & 1.3 (-9)  & 8.0 (-9)  & 6.0 (-10) -- 2.0 (-9)$^6$ \\
CH$_3$CN      & 1.0 (-9)  & 1.5 (-8)  & 6.0 (-10)$^9$ \\
NH$_3$        & 3.4 (-9)  & 1.0 (-5)  & 2.0 (-8)$^{1,4}$ \\
NH$_2$D       & 4.4 (-11) & 2.4 (-7)  & 2.2 (-10)$^2$ \\
NHD$_2$       & 5.6 (-13) & 9.2 (-10) & \\
ND$_3$        & 9.2 (-14) & 1.4 (-11) & \\
N$_2$H$^+$    & 2.8 (-10) & \nodata   & 4.0 (-10)$^1$ \\
N$_2$D$^+$    & 2.2 (-11) & \nodata   & 3.2 (-11)$^2$  \\
C             & 7.8 (-7)  & \nodata   & \\
C$^+$         & 1.8 (-8)  & \nodata   & \\
CH            & 5.2 (-9)  & \nodata   & 2.0 (-8)$^1$ \\
C$_2$H        & 1.2 (-9)  & \nodata   & 2.0 (-8)$^7$ \\
C$_2$D        & 3.5 (-11) & \nodata   & 2.0 (-10)$^8$ \\
C$_3$H        & 5.9 (-7)  & \nodata   & 1.0 (-8)$^7$ \\
C$_2$H$_2$    & 1.7 (-8)  & 1.9 (-6)  & \\
CH$_4$        & 2.7 (-7)  & 1.4 (-6)  & \\
CH$_3$D       & 1.7 (-8)  & 9.2 (-8)  & \\
CH$_2$D$_2$   & 9.3 (-10) & 2.6 (-9)  & \\
CHD$_3$       & 3.3 (-11) & 1.8 (-10) & \\
CD$_4$        & 1.2 (-13) & 6.4 (-13) & \\
\enddata
\tablerefs{{1} \citet{ohisi92}, {2} \citet{tine00}, {3} \citet{turner01},
{4} \citet{pratap97}, {5} \citet{guelin82}, {6} \citet{howe94}, 
{7} \citet{tht00}, {8} \citet{millar89}, {9} \cite{ok98} }
\end{deluxetable}


\begin{deluxetable}{lllll}
\tabletypesize\footnotesize
\tablecolumns{5}
\tablewidth{0pt}
\tablecaption{\label{tab:cd}The column densities, N($x$), calculated at $R$ = 250 AU and a time of 1 Myrs
($a (b)$ represents $a$ $\times$ 10$^b$ cm$^{-2}$).
All models include thermal desorption and this is the only
desorption process included in Model A.
Model B also includes CRH, Model C includes CRH and photodesorption
and  Model D only includes photodesorption.
The column density of H$_2$ = 4.8 $\times$ 10$^{22}$ cm$^{-2}$ for all models.
}
\tablehead{
\colhead{Molecule} & \multicolumn{4}{c}{$N(x)$} \\
& \colhead{Model A} &  \colhead{Model B} & \colhead{Model C} & \colhead{Model D}}
\startdata
{\bf Gas species}\\
HD            & 5.2 (17) & 2.9 (17) & 2.8 (17) & 5.1 (17) \\
CO            & 1.2 (17) & 1.2 (17) & 2.9 (17) & 2.7 (17) \\
H$_2$O        & 2.7 (10) & 4.0 (10) & 1.6 (15) & 1.6 (15) \\
HDO           & 1.4 (9)  & 9.3 (9)  & 2.0 (14) & 2.0 (14)  \\
D$_2$O        & 1.5 (7)  & 1.3 (8)  & 7.1 (12) & 7.0 (12) \\
CO$_2$        & 9.8 (9)  & 1.0 (10) & 1.1 (14) & 1.1 (14) \\
CH$_4$        & 2.9 (14) & 3.3 (14) & 2.9 (14) & 2.5 (14) \\
CH$_3$D       & 3.7 (13) & 4.5 (13) & 3.4 (13) & 2.7 (13)\\
CH$_2$D$_2$   & 3.0 (12) & 3.8 (12) & 2.5 (12) & 1.7 (12) \\
CHD$_3$       & 1.1 (11) & 1.8 (11) & 1.1 (11) & 4.6 (10)\\
CD$_4$        & 1.3 (9)  & 6.9 (9)  & 5.9 (9)  & 2.6 (8) \\
HNC           & 2.0 (10) & 3.7 (10) & 3.2 (12) & 3.1 (12) \\
DNC           & 8.9 (7)  & 2.2 (8)  & 8.5 (10) & 8.6 (10) \\
HCN           & 1.7 (10) & 1.8 (10) & 3.3 (13) & 3.3 (13)\\
DCN           & 2.3 (8)  & 8.9 (8)  & 1.2 (12) & 1.2 (12) \\
CN            & 3.7 (11) & 3.8 (11) & 3.6 (13) & 3.6 (13) \\
H$_2$CO       & 3.3 (11) & 3.3 (11) & 3.9 (12) & 3.9 (12) \\
HDCO          & 6.7 (9)  & 7.5 (9)  & 8.6 (10) & 8.5 (10) \\
D$_2$CO       & 6.0 (8)  & 6.5 (8)  & 1.2 (9)  & 1.2 (9) \\
HC$_3$N       & 1.5 (9)  & 1.6 (9)  & 7.0 (11) & 7.0 (11) \\
DC$_3$N       & 4.6 (7)  & 4.9 (7)  & 2.1 (10) & 2.2 (10) \\
N$_2$         & 3.8 (15) & 3.9 (15) & 2.2 (16) & 2.2 (16) \\
\chem{N_2H^+} & 6.3 (9)  & 2.3 (10) & 2.8 (10) & 1.1 (10)\\
\chem{N_2D^+} & 3.2 (9)  & 1.6 (11) & 1.6 (11) & 3.8 (9) \\
CH$_3$CN      & 2.1 (7)  & 2.1 (7)  & 4.9 (10) & 4.8 (10) \\
CH$_3$OH      & 6.4 (5)  & 6.5 (5)  & 1.5 (11) & 1.5 (11) \\
CH$_3$OD      & 7.9 (3)  & 1.2 (4)  & 1.2 (10) & 1.2 (10) \\
CH$_2$DOH     & 3.3 (4)  & 3.5 (4)  & 1.3 (10) & 1.3 (10) \\
\chem{H_3^+}  & 2.8 (12) & 3.1 (12) & 8.9 (11) & 5.3 (11) \\
\chem{H_2D^+} & 5.3 (11) & 8.2 (11) & 7.7 (11) & 4.8 (11) \\
\chem{HD_2^+} & 6.6 (11) & 1.1 (12) & 1.1 (12) & 6.6 (11) \\
\chem{D_3^+}  & 2.6 (13) & 1.3 (13) & 1.3 (13) & 2.6 (13) \\
HCO$^+$       & 2.9 (12) & 3.3 (12) & 3.2 (12) & 2.8 (12) \\
DCO$^+$       & 3.4 (11) & 3.0 (12) & 3.2 (12) & 4.8 (11) \\
CH$_3^+$      & 9.2 (10) & 9.2 (10) & 1.1 (11) & 1.1 (11) \\
CH$_2$D$^+$   & 1.0 (9)  & 1.0 (9)  & 4.2 (8)  & 4.0 (8) \\
CHD$_2^+$     & 4.4 (7)  & 4.3 (7)  & 1.0 (7)  & 9.1 (6) \\
CD$_3^+$      & 9.6 (5)  & 9.8 (5)  & 1.7 (5)  & 1.6 (5) \\
C$_2$H        & 3.6 (12) & 3.6 (12) & 7.2 (12) & 7.2 (12) \\
C$_2$D        & 8.0 (10) & 8.1 (10) & 2.2 (11) & 2.1 (11)\\
NH$_3$        & 2.7 (11) & 2.9 (11) & 1.4 (13) & 1.4 (13) \\
NH$_2$D       & 3.5 (9)  & 1.9 (10) & 1.3 (12) & 1.2 (12) \\
NHD$_2$       & 2.5 (8)  & 1.5 (9)  & 2.6 (10) & 2.3 (10) \\
ND$_3$        & 1.6 (8)  & 4.9 (8)  & 4.2 (9)  & 3.2 (9)  \\
H             & 3.8 (19) & 4.5 (19) & 4.8 (19) & 4.8 (19) \\
D             & 6.6 (15) & 6.5 (15) & 1.3 (16) & 1.3 (16) \\
\\
{\bf Mantle species} \\
CO          & 4.1 (18) & 4.0 (18) & 4.0 (18) & 4.1 (18) \\
H$_2$O      & 1.2 (19) & 1.2 (19) & 1.2 (19) & 1.2 (19) \\
HDO         & 2.7 (17) & 3.2 (17) & 3.3 (17) & 2.8 (17) \\
D$_2$O      & 4.0 (15) & 1.9 (16) & 1.9 (16) & 4.1 (15) \\
CH$_4$      & 2.5 (17) & 2.7 (17) & 2.7 (17) & 2.4 (17)  \\
CH$_3$D     & 2.2 (16) & 5.5 (16) & 5.4 (16) & 2.1 (16) \\
CH$_2$D$_2$ & 1.9 (15) & 1.1 (16) & 1.1 (16) & 1.8 (15)\\
HCN         & 4.5 (17) & 4.9 (17) & 4.7 (17) & 4.7 (17) \\
DCN         & 9.5 (15) & 8.9 (15) & 8.3 (15) & 8.9 (15) \\
H$_2$CO     & 3.9 (17) & 3.9 (17) & 3.8 (17) & 3.7 (17) \\
HDCO        & 3.3 (15) & 3.8 (15) & 3.6 (15) & 3.0 (15) \\
D$_2$CO     & 4.1 (13) & 4.2 (13) & 3.5 (13) & 3.6 (13) \\
NH$_3$      & 1.0 (18) & 1.0 (18) & 1.0 (18) & 9.6 (17) \\
NH$_2$D     & 3.4 (16) & 2.3 (17) & 2.3 (17) & 3.3 (16) \\
NHD$_2$     & 2.1 (15) & 5.2 (16) & 5.2 (16) & 2.0 (15) \\
CH$_3$OH    & 4.3 (15) & 4.3 (15) & 3.9 (15) & 3.9 (15) \\
CH$_2$DOH   & 3.1 (14) & 3.2 (14) & 2.1 (14) & 2.0 (14) \\
CH$_3$OD    & 1.7 (13) & 1.7 (13) & 2.4 (13) & 2.4 (13) \\
N$_2$       & 1.9 (17) & 3.2 (16) & 3.2 (16) & 1.9 (17) \\
\enddata
\end{deluxetable}

\begin{deluxetable}{llllllll}
\tabletypesize\footnotesize
\tablecolumns{8}
\tablewidth{0pt}
\tablecaption{\label{tab:dh}The molecular D/H ratios
calculated in the cloud model and in the disk.  For
the disk we present two ratios (1) the midplane value
and (2) N(XH)/N(XD), where XH and XD are the non-deuterated
and deuterated forms of the molecule respectively. The ratios
are given in the form $a (-b)$ = $a$ $\times$ 10$^{-b}$.
\nodata indicates that the molecular gas phase abundances are too low to determine meaningful
D/H ratios. The midplane ratios in Models B and C are similar 
since the photodesorption included in the latter model is not efficient
in the midplane.  In Models A and D, the lack of CRH means that most species are
frozen out in the midplane and this high depletion results in
high deuteration of H$_3^+$.}
\tablehead{
\colhead{Molecule} & \colhead{Cloud} & \multicolumn{2}{c}{Midplane} & \multicolumn{4}{c}{N(XH)/N(XD)}\\
& & \colhead{Models A \& D} & \colhead{(Models B \& C)} & \colhead{Model A} & \colhead{Model B} & \colhead{Model C} & \colhead{Model D}\\
}
\startdata
\sidehead{Gas Species}
H$_2$D$^+$/H$_3^+$    & 1.6 (-1) & 9.9 (-1)& 9.5 (-1) & 1.9 (-1) & 2.6 (-1) & 8.7 (-1) & 9.1 (-1) \\
HD$_2^+$/H$_3^+$      & 2.4 (-2) & 1.6     & 1.5      & 2.4 (-1) & 3.5 (-1) & 1.2 (-1) & 1.2 (-1) \\
D$_3^+$/H$_3^+$       & 3.1 (-3) & 76.9    & 25.1     & 9.3      & 4.1      & 14.6     & 49.1 \\
CH$_2$D$^+$/CH$_3^+$  & 8.7 (-2) & \nodata & \nodata  & 1.1 (-2) & 1.1 (-2) & 3.8 (-3) & 3.6 (-3) \\
CHD$_2^+$/CH$_3^+$    & 3.7 (-3) & \nodata & \nodata  & 4.8 (-4) & 4.7 (-4) & 9.1 (-5) & 8.3 (-5) \\
CD$_3^+$/CH$_3^+$     & 6.8 (-5) & \nodata & \nodata  & 1.0 (-5) & 1.1 (-5) & 1.5 (-6) & 1.5 (-6)\\
DCO$^+$/HCO$^+$       & 7.2 (-2) & \nodata & 12.1     & 1.2 (-1) & 9.1 (-1) & 1.0      & 1.7 (-1) \\
DCN/HCN               & 3.3 (-2) & \nodata & \nodata  & 1.4 (-2) & 4.9 (-2) & 3.6 (-2) & 3.6 (-2) \\
DNC/HNC               & 1.2 (-2) & \nodata & \nodata  & 4.5 (-3) & 5.9 (-3) & 2.7 (-2) & 2.8 (-2) \\
HDCO/H$_2$CO          & 4.2 (-2) & \nodata & \nodata  & 2.0 (-2) & 2.3 (-2) & 2.2 (-2) & 2.2 (-2) \\
D$_2$CO/H$_2$CO       & 1.0 (-3) & \nodata & \nodata  & 1.8 (-3) & 2.0 (-3) & 3.1 (-4) & 3.1 (-4) \\
NH$_2$D/NH$_3$        & 1.3 (-2) & \nodata & \nodata  & 1.3 (-2) & 6.6 (-1) & 9.3 (-2) & 8.6 (-2) \\
NHD$_2$/NH$_3$        & 1.6 (-4) & \nodata & \nodata  & 9.3 (-4) & 5.2 (-3) & 1.9 (-3) & 1.6 (-3) \\
ND$_3$/NH$_3$         & 2.7 (-5) & \nodata & \nodata  & 5.9 (-4) & 1.7 (-3) & 3.0 (-4) & 2.3 (-4) \\
HDO/H$_2$O            & 4.4 (-2) & \nodata & \nodata  & 5.2 (-2) & 2.3 (-1) & 1.3 (-1) & 1.3 (-1) \\
D$_2$O/H$_2$O         & 2.0 (-4) & \nodata & \nodata  & 5.6 (-4) & 3.3 (-3) & 4.4 (-3) & 4.4 (-3) \\
CH$_3$D/CH$_4$        & 6.3 (-2) & \nodata & \nodata  & 1.3 (-1) & 1.4 (-1) & 1.2 (-1) & 1.1 (-1) \\
CH$_2$D$_2$/CH$_4$    & 3.4 (-3) & \nodata & \nodata  & 1.0 (-2) & 1.2 (-2) & 8.6 (-3) & 6.8 (-3) \\
CHD$_3$/CH$_4$        & 1.2 (-4) & \nodata & \nodata  & 3.8 (-4) & 5.5 (-4) & 3.8 (-4) & 1.8 (-4) \\
CD$_4$/CH$_4$         & 4.4 (-7) & \nodata & \nodata  & 4.5 (-6) & 2.1 (-5) & 2.0 (-5) & 1.0 (-6) \\
N$_2$D$^+$/N$_2$H$^+$ & 7.9 (-2) & \nodata & 12.0     & 5.1 (-1) & 7.0      & 5.7      & 3.5 (-1) \\
CH$_3$OD/CH$_3$OH     & 2.5 (-2) & \nodata & \nodata  & 1.2 (-2) & 1.8 (-2) & 8.0 (-2) & 8.0 (-2) \\
CH$_2$DOH/CH$_3$OH    & 7.3 (-2) & \nodata & \nodata  & 5.1 (-2) & 5.4 (-2) & 8.7 (-2) & 8.7 (-2) \\
D/H                   & 2.1 (-2) & 1.5     & 1.0      & 1.7 (-4) & 1.4 (-4) & 2.7 (-4) & 2.7 (-4) \\
\\
\sidehead{Mantle species}\\
HDO/H$_2$O         & 2.0 (-2) & 2.1 (-2) & 2.4 (-2) & 2.3 (-2) & 2.7 (-2) & 2.8 (-2) & 2.3 (-2) \\
D$_2$O/H$_2$O      & 1.5 (-4) & 1.8 (-4) & 1.3 (-3) & 3.3 (-4) & 1.6 (-3) & 1.6 (-3) & 3.4 (-4) \\
NH$_2$D/NH$_3$     & 2.4 (-2) & 2.4 (-2) & 2.3 (-1) & 3.4 (-2) & 2.3 (-1) & 2.3 (-1) & 3.4 (-2) \\
NHD$_2$/NH$_3$     & 9.2 (-5) & 1.2 (-4) & 5.5 (-2) & 2.1 (-3) & 5.2 (-2) & 5.2 (-2) & 2.1 (-3)\\
CH$_3$D/CH$_4$     & 6.6 (-2) & 6.7 (-2) & 2.1 (-1) & 8.8 (-2) & 2.0 (-1) & 2.0 (-1) & 8.8 (-2) \\
CH$_2$D$_2$/CH$_4$ & 1.9 (-3) & 3.8 (-3) & 4.6 (-2) & 7.6 (-3) & 4.1 (-2) & 4.1 (-2) & 7.5 (-3) \\
DCN/HCN            & 6.1 (-3) & 2.0 (-2) & 1.9 (-2) & 2.1 (-2) & 1.8 (-2) & 1.8 (-2) & 1.9 (-2)  \\
\enddata
\end{deluxetable}

\begin{deluxetable}{llllllll}
\tabletypesize\footnotesize
\tablecolumns{8}
\tablewidth{0pt}
\tablecaption{\label{tab:obs}A comparison of the results of Models B, C and D with the
observations.  The calculated column densities (cm$^{-2}$) are given at $t$ = 1 Myrs and $R$ = 250 AU. 
$a (b)$ represents $a$ $\times$ 10$^b$. 
For LkCa15 the column densities calculated from the single dish 
observations assume that the disk has a radius of 450 AU.  For
TW Hya the disk is assumed to be 165 AU in radius.  For each model N(H$_2$) = 4.2 \x 10$^{22}$ cm$^{-2}$.
}
\tablehead{
\colhead{Molecule} & \colhead{Model B} & \colhead{Model C} & \colhead{Model D} & \colhead{DM Tau$^1$} & \multicolumn{2}{c}{LkCa15} & \colhead{TW Hya$^3$}\\
\colhead{} & \colhead{} & \colhead{} & \colhead{} & \colhead{} & \colhead{Interferometer$^2$} & \colhead{Single dish$^3$} 
}
\startdata
CO            & 1.2 (17) & 2.9 (17) & 2.7 (17) & 5.7 (16)       & 1.7 (18)$^7$       & 1.9 (16)       & 3.2 (16) \\
HCN           & 1.8 (10) & 3.3 (13) & 3.3 (13) & 2.1 (12)       & 2.4 (13)           & 1.8 (12)       & 8.5 (12)$^6$ \\
HNC           & 3.7 (10) & 3.2 (12) & 3.1 (12) & 9.1 (11)       & $<$ 5.4 (12)       & \nodata        & $<$ 1.4 (12)\\
CN            & 3.8 (11) & 3.6 (13) & 3.6 (13) & 9.5 - 12 (12)  & 9.7 -- 25 (13)     & 1.5 (13)       & 6.6 (13) \\
CH$_3$OH      & 6.5 (5)  & 1.5 (11) & 1.5 (11) & \nodata        & 7.3 -- 18 (14)     & $<$ 7.1 (13)   & $<$ 1.1 (13)\\
H$_2$CO       & 3.3 (11) & 3.9 (12) & 3.9 (12) & 7.6 -- 19 (11) & 7.2 -- 19 (12)$^4$ & 7.1 -- 51 (11) & $<$ 8.0 (11) \\
HCO$^+$       & 3.3 (12) & 3.2 (12) & 2.8 (12) & 4.6 -- 28 (11) & 1.5 (13)           & 3.3 (11)       & 8.5 (12)$^6$ \\
C$_2$H        & 3.6 (12) & 7.2 (12) & 7.2 (12) & 4.2 (13)       & \nodata            &  \nodata       & \nodata \\
N$_2$H$^+$    & 2.3 (10) & 2.8 (10) & 1.1 (10) & $<$ 7.6 (11)   & 3.1 (13)$^7$       & $<$ 1.4 (12)   & $<$ 1.0 (13) \\
DCO$^+$       & 3.0 (12) & 3.2 (12) & 4.8 (11) & \nodata        & \nodata            & $<$ 2.9 (11)   & 3.0 (11)$^6$ \\
DCN           & 8.9 (8)  & 1.2 (12) & 1.2 (12) & \nodata        & \nodata            & 1 (13)$^8$     & $<$ 4.0 (10) \\
HDO           & 9.3 (9)  & 2.0 (14) & 2.0 (14) & 1.6 (13)       & \nodata            & 2 -- 7 (14)$^8$&               \\
\chem{H_2D^+} & 8.2 (11) & 7.7 (11) & 4.8 (11) & 8.8 (12)$^5$   & \nodata            & $<$ 2.0 (14)$^5$ &  1.3 (14)     \\
\enddata
\tablerefs{(1) \citet{aikawa02} (derived from \citet{dutrey97}), (2) \citet{qi01}, (3) \citet{thi04},
(4) \citet{aikawa03}, (5) \citet{cec05}, (6) \citet{vd03}, (7) \citet{qi03}, (8) \citet{qi01}}
\end{deluxetable}

\clearpage

\end{document}